\documentclass[prd,preprint,nofootinbib]{revtex4-1}
\usepackage{graphicx}
\usepackage{epsfig}
\usepackage{graphics}
\usepackage{amsmath}
\usepackage{latexsym}
\usepackage{dcolumn}
\usepackage{bm}
\usepackage{algorithmicx}
\usepackage[pdftex]{color}
\usepackage[utf8]{inputenc}
\usepackage{booktabs}
\usepackage{multirow}
\usepackage{siunitx}
\begin{document}

\title{Vectorial and spinorial perturbations in Galileon Black Holes: Quasinormal modes, quasiresonant modes and stability}
%%%%%%%%%%%%%%%%%%%%%%%%%%%%%%%%%%%%%%%%%%%%%%%%%%%
\author{E. Abdalla}
\email{eabdalla@usp.br}
\affiliation{Instituto de F\'{i}sica, Universidade de S\~ao Paulo, Caixa Postal 66318, 05314-970, S\~ao Paulo, São Paulo , Brazil}

\author{B. Cuadros-Melgar}
\email{bertha@usp.br}
\affiliation{Escola de Engenharia de Lorena, Universidade de S\~ao Paulo, Estrada Municipal do Campinho S/N, Bairro Campinho,
12602-810, Lorena, São Paulo, Brazil}

\author{Jeferson de Oliveira}
\email{jeferson@fisica.ufmt.br}
\affiliation{Instituto de F\'{i}sica, Universidade Federal de Mato Grosso, 78060-900, Cuiab\'a, Mato Grosso, Brazil}

\author{A. B. Pavan}
\email{alan@unifei.edu.br}
\affiliation{Instituto de F\'{i}sica e Qu\'{i}mica, Universidade Federal de Itajub\'{a}, Caixa Postal 50, 37500-903, Itajub\'{a}, Minas Gerais, Brazil}
\author{C. E. Pellicer}
\email{carlos.pellicer@ect.ufrn.br}
\affiliation{Escola de Ci\^encias e Tecnologia, Universidade Federal do Rio Grande do Norte, Caixa Postal 1524, 59072-970, Natal, Rio Grande do Norte, Brazil}

%%%%%%%%%%%%%%%%%%%%%%%%%%%%%%%%%%%%%%%%%%%%%%%%%%%
\begin{abstract}
In this work we have considered a model that includes the interaction
of gravity and matter fields with Galilean invariance (the so-called
derivative coupling) as well as some
corresponding black hole type solutions. Quasinormal perturbations of
two kinds of matter fields have been computed by different methods.
The effect of the derivative coupling in the quasinormal spectrum has been analyzed and evaluated.

\end{abstract}
%%%%%%%%%%%%%%%%%%%%%%%%%%%%%%%%%%%%%%%%%%%%%%%%%%%

%04.30.Nk Wave propagation and interactions
%04.20.Ex Initial value problem, existence and uniqueness of solutions

\maketitle

%%%%%%%%%%%%%%%%%%%%%%%%%%%%%%%%%%%%%%%%%%%%%%%%%%%
\section{Introduction}
%%%%%%%%%%%%%%%%%%%%%%%%%%%%%%%%%%%%%%%%%%%%%%%%%%%
%%%%%%%%%%%%%%%%%%%%%%%%%%%%%%%%%%%%%%%%%%%%%%%%%%%%%%%%%%%%%%%%%%%%%%%%%%%%%%%%%%%%%%%%%%%%%%%%%%%%%%
%%%%%%%%%%%%%%%%%%%%%%%%%%%%%%%%%%%%%%%%%%%%%%%%%%%%%%%%%%%%%%%%%%%%%%%%%%%%%%%%%%%%%%%%%%%%%%%%%%%%%%1234567

Higher order terms in a field theory action including gravity are expected to appear in
view of the high nonlinearity of gravity as well as due to corrections from
string theories.  Such higher order terms are rather unwanted, especially if
there are higher order derivatives in the equations of motion, a case leading
to a Hilbert space with a nonpositive scalar product. Even at a classical level
higher derivative interactions are well known to lead to instabilities.
However, Horndeski showed that there is a large class of fields which,
in spite of having  derivative terms of arbitrary order in the action, yield equations of motion at most second order in the derivatives~\cite{horndeski}, preventing, in principle, instabilities.

More recently, these ideas were used to describe a scalar fulfilling a
second order equation of motion and which, moreover, obeyed a Galilean
invariance~\cite{nicolis,galilean_invariance2}. Cosmology also has 
several implications in the case of the presence of Horndeski
scalars~\cite{Koutsoumbas_et_al_2018}. Problems related to
instabilities are very important  and Horndeski theories offer a good
example of such phenomena~\cite{Kolyvaris_etal_2018}. The most
important new term in the action is the coupling of the scalar field
derivative with the Einstein curvature tensor $ G_{\mu\nu}$, the
so-called derivative coupling term,
\begin{equation}
\delta {\cal L}= \bar z \sqrt{-g}\, G^{\mu\nu}\partial_\mu\phi\partial_\nu\phi\quad .
\end{equation}
The new interaction term behaves as a friction term for $\bar z > 0$
in cosmological contexts, while a negative sign may lead to instability.
Moreover, the derivative coupling term has been used in
other physical contexts with interesting results, such as new
solutions and nonperturbative effects~\cite{rinaldi, Anabalon:2013oea,Cisterna:2014nua,Cisterna:2015yla, Cisterna:2016vdx,Brihaye:2016lin,minamitsujisol}, quasinormal
modes~\cite{minamitsuji,konoplya}, structure formation~\cite{bellini_sawicki},
self-accelerating solutions~\cite{nicolis}, and disformal
transformations in the dark sector~\cite{zumalacarregui_koivisto_mota}. Such a term also represents a drag or a boost to the fields, depending on the sign; thus, it can be of importance to the AdS/CFT conjecture as a means of a possible source of new physics, since the coupling to the Einstein tensor can be related to impurities in a superconductor~\cite{papaandkuang}.

Our primary aim is to consider the effects of such a term on the
stability of a specific black hole solution. The probe fields we will
consider are vectors and spinors obeying Maxwell/Proca and Dirac
equations, respectively. As most studies of quasinormal modes around
black holes focus on scalar fields due to their applications in
cosmology, our motivation here is to use other fields with richer
structure and more degrees of freedom that can reveal new features
of the background model. In the following sections, we describe the Galileon
black hole metric considered here, set the corresponding perturbation
equations, and compute as well as analyze the quasinormal spectrum
using numerical methods.

%%%%%%%%%%%%%%%%%%%%%%%%%%%%%%%%%%%%%%%%%%%%%%%%%%%%%%%%%%%%%%%%%%%%%%%%%%%%%%%%%%%%%%%%%%%
%%%%%%%%%%%%%%%%%%%%%%%%%%%%%%%%%%%%%%%%%%%%%%%%%%%%%%%%%%%%%%%%%%%%%%%%%%%%%%%%%%%%%%%%%%%%%

\section{Galileon Black Hole Solutions}

We consider a model described by an action consisting of the Einstein-Hilbert term plus a scalar field kinetically coupled to the curvature given by
\begin{equation}\label{action}
S=\frac{1}{2}\int d^4 x \sqrt{-g} \left[\beta R - \left(g^{\mu\nu} - \bar z G^{\mu\nu}\right)\partial_\mu \phi \partial_\nu \phi\right]\,,
\end{equation}
where $\beta=m_P ^2$, $\bar z=\frac{z}{m_P ^2}$ and $G^{\mu\nu}$ is
the Einstein tensor. The presence of this nonminimal derivative
coupling has far-reaching consequences. One of the most important is
the fact that if it plays the role of dark energy the speed of
propagation of gravitational waves get corrections that may be easily
detected but have not been until present
times~\cite{LIGOCollaboration2016,GongPapantonopoulos2017}, which has a
negative impact on its uses for cosmology \footnote{Nonetheless, with some
combinations of Horndeski Lagrangians it is still possible to
obtain $c_{gw}\approx c$ provided that their effects are negligible at
late times~\cite{GongPapantonopoulos2017}}. However, in view of the potential ubiquity of derivative terms in string inspired cosmology, Horndeski theories remain important as a realistic possibility, and the understanding of its role (and possible outcome of the above failure) is worth considering, at least for very high energies.

Since there is no scalar potential, the action is invariant under shift symmetry $\phi \rightarrow \phi + const$. This is precisely the reason to name $\phi$ a Galileon field. Moreover, the $\bar z$ term plays the role of the friction term alluded to above.

Varying the action with respect to the metric, we obtain the corresponding field equation,
\begin{eqnarray}\label{field}\nonumber
&&\beta \left(\frac{1}{2} g^{\lambda\kappa}R-R^{\lambda\kappa}\right)
  +  \frac{1}{2}\partial^{\nu}\phi \partial_\nu \phi
  g^{\lambda\kappa}-\partial^\lambda \phi \partial^\kappa \phi + \bar z \left[\frac{1}{2} R^{\mu\nu} \partial_\mu \phi \partial_\nu \phi g^{\lambda\kappa} - 2 R^{\mu\kappa}\partial_\mu \phi\partial^\lambda \phi
  + \frac{1}{2} R\partial^\lambda\phi \partial^\kappa \phi\right. \\ \nonumber
&+&\left. \frac{1}{2}\left(R^{\lambda\kappa}-\frac{1}{2}g^{\lambda\kappa}R\right) \partial^\nu \phi \partial_\nu \phi - \partial^\lambda \partial^\kappa \phi \Box \phi - \frac{1}{2} \partial_\alpha \partial_\nu \phi \partial^\alpha \partial^\nu \phi g^{\lambda\kappa} + \partial^\lambda\partial_\nu \phi \partial^\kappa \partial^\nu \phi + \frac{1}{2}(\Box \phi)^2 g^{\lambda\kappa}\right] =0 .\\
\end{eqnarray}
Furthermore, varying Eq.(\ref{action}) with respect to $\phi$, we obtain the Galileon field equation,
\begin{equation}\label{escalar}
\partial_\mu \left[\sqrt{-g} \left(\partial^\mu \phi - 2 \bar z
  G^{\mu\nu} \partial_\nu \phi \right)\right]=0\,.
\end{equation}

In the spirit of Refs. \cite{Amendola,Sushkov} black hole solutions for $z>0$(for the case $z<0$ see coment \footnote{In the case $z<0$ the solution tends to a dS space. However, the metric turns out to be non-differentiable and a non-trivial stress-energy tensor needs to be added to the Lagrangian.}), as those (\cite{rinaldi,minamitsujisol}) can be obtained in the standard way from (\ref{field}) and (\ref{escalar}). With the ansatz
\begin{equation}\label{metric}
ds^2=-F(r) dt^2 + H(r) dr^2 +r^2 (d\theta^2 + \sin^2\theta d\phi^2)\,,
\end{equation}
one finds, as a result,
\begin{eqnarray}
F(r) &=& \frac{3}{4} + \frac{r^2}{L^2} - \frac{2M}{m_P ^2r} +
\frac{\sqrt{\bar z}}{4r}\arctan\left(\frac{r}{\sqrt{\bar z}}\right)\,, \label{Fg}\\
H(r) &=& \frac{(r^2 +2\bar z)^2}{4(r^2+\bar z)^2 F(r)}\,, \label{Hg}\\
\left[\phi'(r)\right]^2 &=& -\frac{m_P ^2r^2(r^2+2\bar z)^2}{4\bar
  z(r^2+\bar z)^3F(r)}\,,
\end{eqnarray}
where $L^2=12\bar z$ and $M$  is an integration constant related to
the black hole mass. We notice that $\bar z$ behaves as a
nonperturbative parameter that interpolates between the Schwarzschild
solution (for $z \rightarrow \infty$) and Schwarzschild anti-de Sitter (AdS)
solution. Out of these limits the Galileon black hole geometry is asymptotically AdS, which makes it interesting in the context of AdS/CFT correspondence.

A thermodynamical analysis of these solutions shows that large mass or small $z$ parameter black holes are thermodynamically stable, while small mass black holes or alternatively having a large $z$ parameter undergo a phase transition to the vacuum solution~\cite{rinaldi}. As a thermodynamical stability or instability does not imply a dynamical one {\it a priori}, we are interested in studying the evolution of matter fields in these black hole backgrounds with the aim of testing not only its stability but also to understand their interpretation in the case of AdS/CFT correspondence.

%%%%%%%%%%%%%%%%%%%%%%%%%%%%%%%%%%%%%%%%%%%%%%%%%%%%%%%%%%%%%%%%%%%%%%%%%%%%%%%%%%%%%%%%%%%%%%

%%%%%%%%%%%%%%%%%%%%%%%%%%%%%%%%%%%%%%%%%%%%%%%%%%%
\section{Vector Field Perturbations}  \label{electroperturbation}
%%%%%%%%%%%%%%%%%%%%%%%%%%%%%%%%%%%%%%%%%%%%%%%%%

Electromagnetic perturbations are important in the context of the AdS/CFT
conjecture since they can be related to perturbations of generic
supergravity gauge fields. Moreover, Maxwell and Proca fields have
unique features,  possibly with an impact on tera-electron-volt scale gravity scenarios~\cite{NPB,miniblackholes}. In addition, higher order couplings including gauge fields have several new implications in the dynamics of gravity and spacetime; see Refs. \cite{ChenJingPRD2014} for discussions on this point.

\subsection{Maxwell field}

The evolution of the electromagnetic perturbation is given by the Maxwell equations without source,
\begin{eqnarray}
\label{maxequation}
\nabla_{\nu}F^{\nu\mu}=0\, ,
\end{eqnarray}
where the Maxwell tensor is given by
\begin{eqnarray}
\label{maxtensor}
F_{\mu\nu}=\partial_\mu A_\nu -\partial_\nu A_\mu
\end{eqnarray}

The vector potential can be decomposed in components with odd (axial) and even (polar) parity as
\begin{equation}\label{decomp}
A_{\mu}(t,r,\theta,\phi)=\sum_{\ell,m}\left(\left[\begin{array}{c}
0\\
0 \\
\frac{a(r,t)}{\sin\theta}\ \frac{\partial Y}{\partial \phi}\\
-a(r,t)\ \sin\theta\ \frac{\partial Y}{\partial \theta}
\end{array}\right]+
\left[\begin{array}{c}
f(r,t) Y \\
j(r,t) Y \\
k(r,t)\ \frac{\partial Y}{\partial \theta} \\
k(r,t)\ \frac{\partial Y}{\partial \phi}
\end{array}\right]\right)\,.
\end{equation}

Substituting back into Maxwell equations (\ref{maxequation}) we obtain the  equation of motion
\begin{equation}
\label{electropertaxial}
-\frac{\partial^2 \Psi(r,t)}{\partial t^2} + \frac{\partial^2 \Psi(r,t)}{\partial r_{*}^2}  - V_M(r) \Psi(r,t) = 0\, ,
\end{equation}
where the tortoise coordinate is given by
\begin{equation}\label{tartaruga1}
dr_*=\sqrt{\frac{H(r)}{F(r)}}dr\,,
\end{equation}
and the wave function $\Psi(r,t)$ is a linear combination of $a(r,t)$,
$f(r,t)$, $j(r,t)$, and $k(r,t)$ as follows
\begin{equation}\label{axpol}
\Psi^{axial} (r,t)= a(r,t)\,, \qquad \Psi^{polar}(r,t) = \frac{r^2 [j(r,t)_{,t}-f(r,t)_{,r}]}{\sqrt{F(r)H(r)}\ \ell(\ell+1)}\,.
\end{equation}
In both cases the effective potential can be written as
\begin{equation}
\label{electroeffpotaxial}
V_M(r) = F(r)\ \frac{\ell(\ell+1)}{r^2}\,.
\end{equation}
Inspecting Eqs.(\ref{electropertaxial}) and (\ref{electroeffpotaxial})
one can see that the electromagnetic perturbations have a simplifying
symmetry obtained by rescaling the spacetime variables
$t=\tau\sqrt{\bar z}$ and $r=x\sqrt{\bar z}$ and the black hole mass
as $M=M^\prime \sqrt{\bar z}$.
Such a result can be explicitly checked in the corresponding tables
shown in the Appendix.
%%%%%%%%%%%%%%%%%%%%%%%%%%%%%%%%%%%%%%%%%%FIGURA 1 %%%%%%%%%%%%%%%%%%%%%%%
\begin{figure}[h!]
\begin{eqnarray}
\rotatebox{0}
{\includegraphics[width=.5\textwidth]{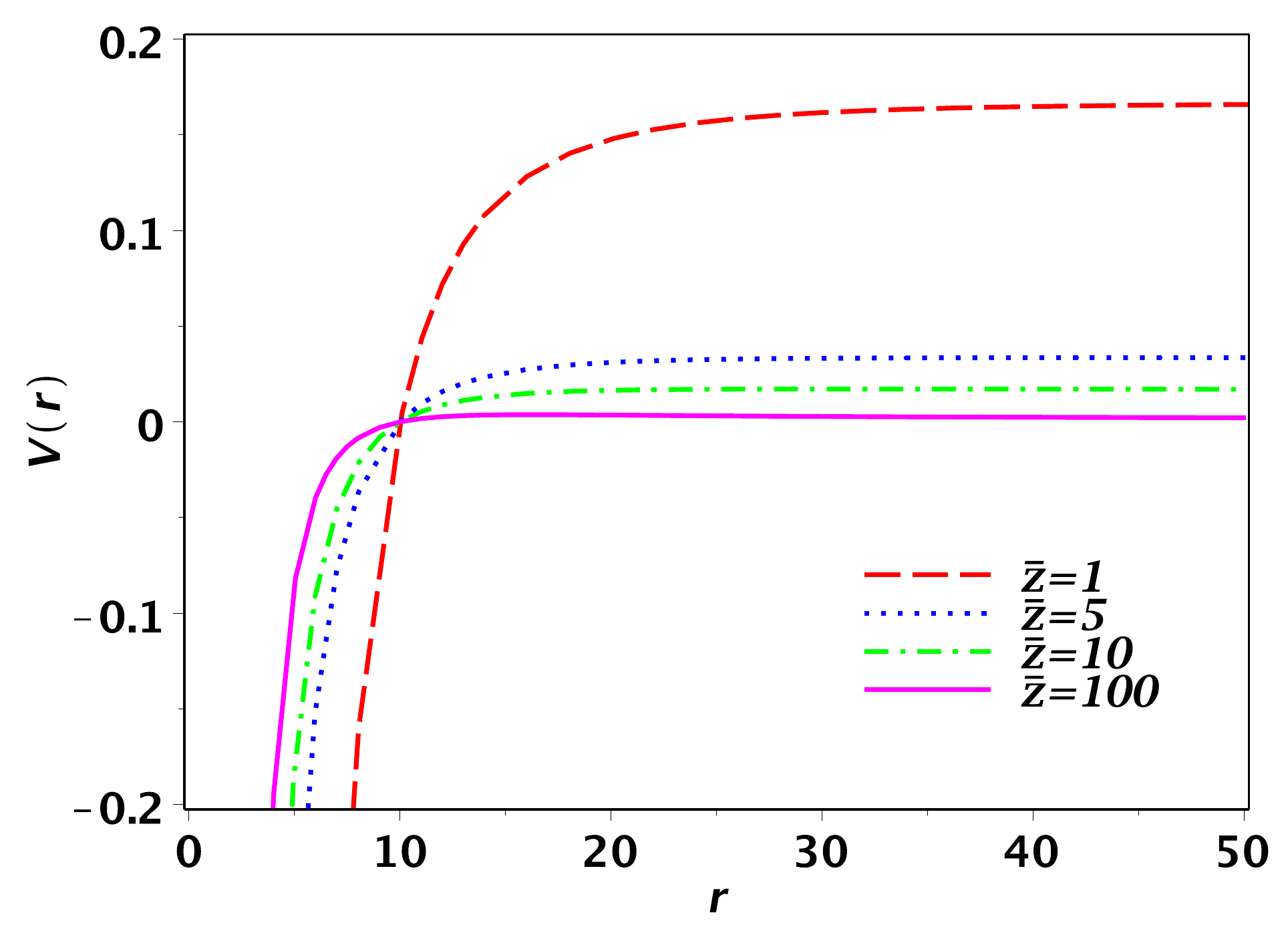}}
\rotatebox{0}
{\includegraphics[width=.5\textwidth]{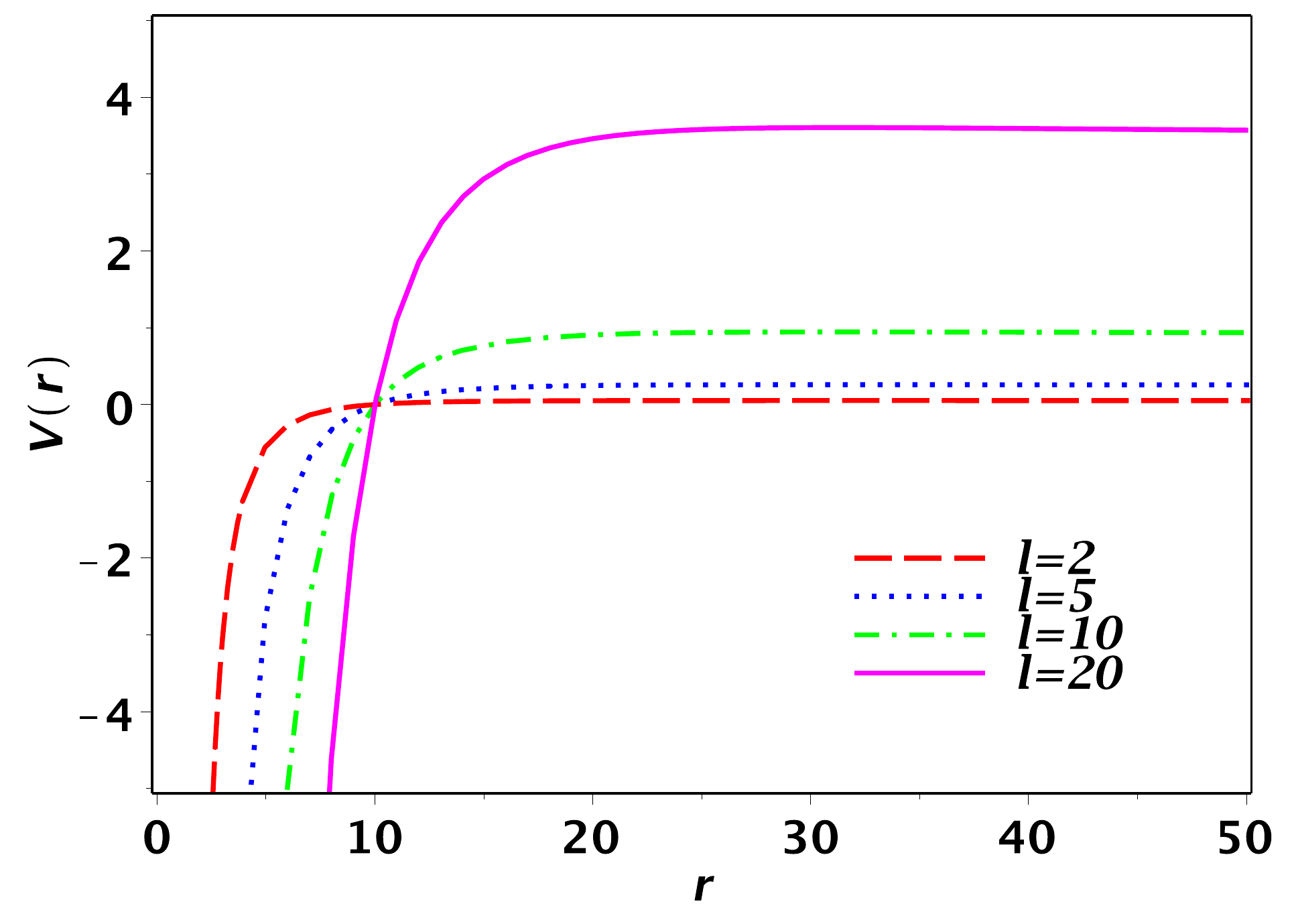}}\nonumber
\end{eqnarray}
\caption{Effective potential as a function of $r$ for Maxwell
  perturbations fixing the event horizon at $r_+=10$ for several
  values of $\bar z$ fixing $\ell=1$ (left) and several values of
  $\ell$ fixing $\bar z=10$.}
\label{potmax}
\end{figure}
%%%%%%%%%%%%%%%%%%%%%%%%%%%%%%%%%%%%%%%%%%%%%%%%%%%%%%%%%%%%%
In the spatial infinity, the electromagnetic effective potential goes to a constant depending on $\bar{z}$ and $\ell$ as
\begin{equation}
\label{electroepotaxialasymp}
V_M(r) \sim \frac{\ell(\ell+1)}{12 \bar{z}}\,.
\end{equation}
Moreover, $\Psi$ becomes independent of $\bar{z}$,
\begin{equation}
\label{electtropsiasymp}
\Psi(r)=C_1+\frac{C_2}{r}\,.
\end{equation}

We plotted the Maxwell potential (\ref{electroeffpotaxial}) as a function
of $r$ in Fig.\ref{potmax}. We can see that as $\bar z$ increases for
fixed multipole number $\ell$ the asymptotic value becomes lower. The inverse
effect is produced by increasing $\ell$ for fixed $\bar z$. These
results are in perfect agreement with Eq.(\ref{electroepotaxialasymp}).

%%%%%%%%%%%%%%%%%%%%%%%%%%%%%%%%%%%%%%%%%%%%%%%%%%%%%%%%%%%%%%%%%%%%%%%%%%%%%%%%%%%%%%%%%%%%%%%%%%%%%%%%%%%%%%%%%%%%%%%%%%%%%%%%%%%%%%%%%%%%%%%%%%%%%%%%%%%%

\subsection{Proca field}
%%%%%%%%%%%%%%%%%%%%%%%%%%%%%%%%%%%%%%%%%%%%%%%%%%%%%%%%%%%%%%%%%%%%%%%%%%%%%%%%%%%%%%%%%%%%%%%%%%%%%%%%%%%%%%%%%%%%%%%%%%%%%%%%%%%%%%%%%%%%%%%%%%%%%%%%%%%%%
Now, we consider massive electromagnetic perturbations, which can be modeled using Proca field equations given by
\begin{eqnarray}
\label{procaequation}
\nabla_{\nu}F^{\nu\mu}-m^2A^{\mu}=0\quad ,
\end{eqnarray}
where $m$ is the mass of the Proca field. The above equation can be decomposed in odd (axial) and even (polar) components as in (\ref{decomp}). The equation of motion for the axial component of the vector field turns out to be
\begin{equation}
\label{procapertaxial}
-\frac{\partial^2 \Psi_P^{axial}}{\partial t^2} + \frac{\partial^2 \Psi_P^{axial}}{\partial r_{*}^2}  - V_P ^{axial} (r) \Psi_P^{axial} = 0\,,
\end{equation}
where again $\Psi_P^{axial}(r,t)$ is shown in (\ref{axpol}), the tortoise coordinate is given by (\ref{tartaruga1}),
and the effective potential $V_P^{axial}$ reads
\begin{equation}
\label{procaeffpotaxial}
V_P^{axial}(r) = F(r)\ \left[\frac{\ell(\ell+1)}{r^2}+m^2\right].
\end{equation}

%%%%%%%%%%%%%%%%%%%%%%%%%%%%%%%%%% FIGURA 2 %%%%%%%%%%%%%%%%%%%%%%%%%%%%%%%%%%%%%%%%%%%
\begin{figure}[h]
\begin{eqnarray}
\rotatebox{0}
{\includegraphics[height=.35\textwidth]{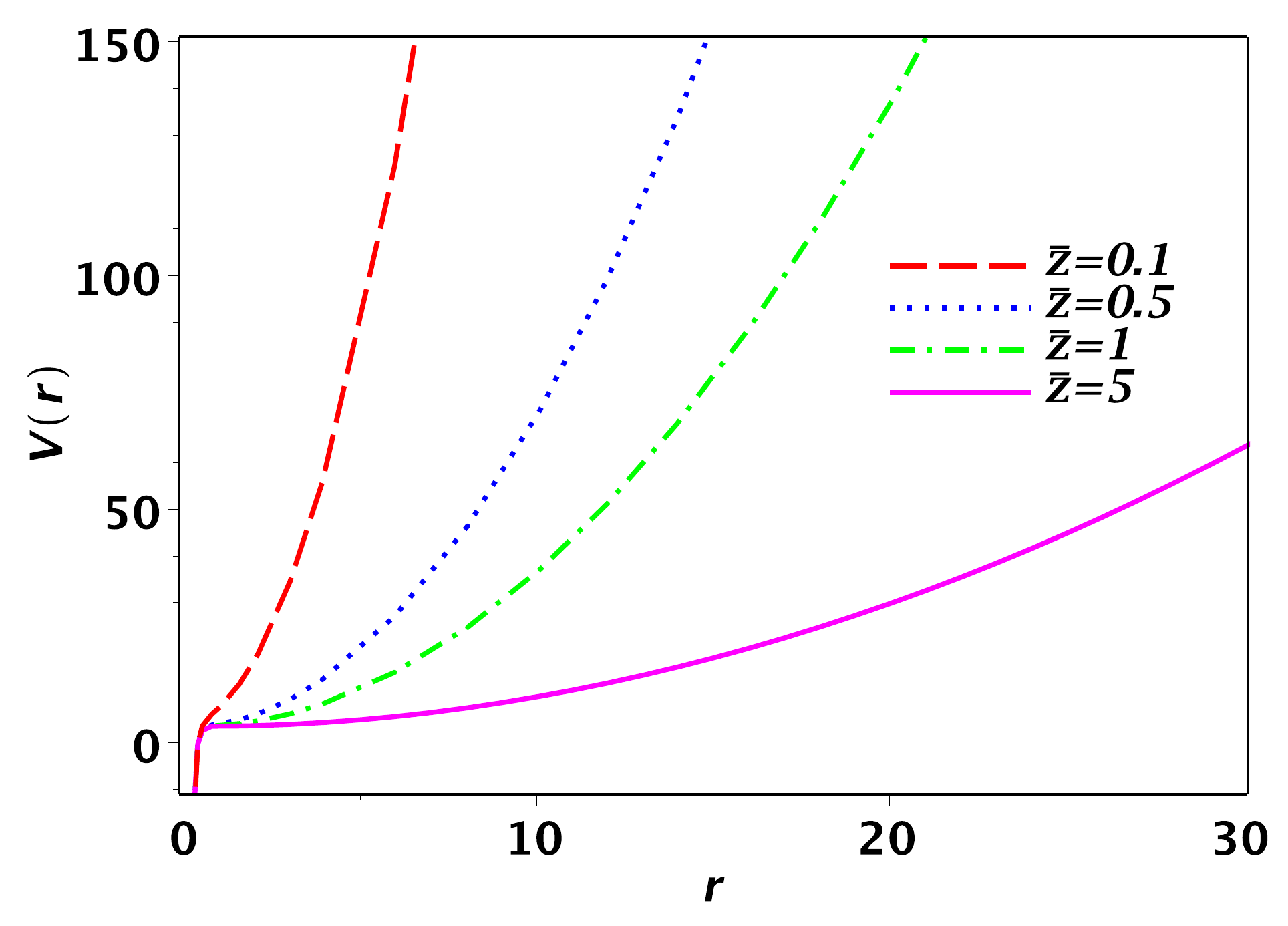}}
\rotatebox{0}
{\includegraphics[height=.35\textwidth]{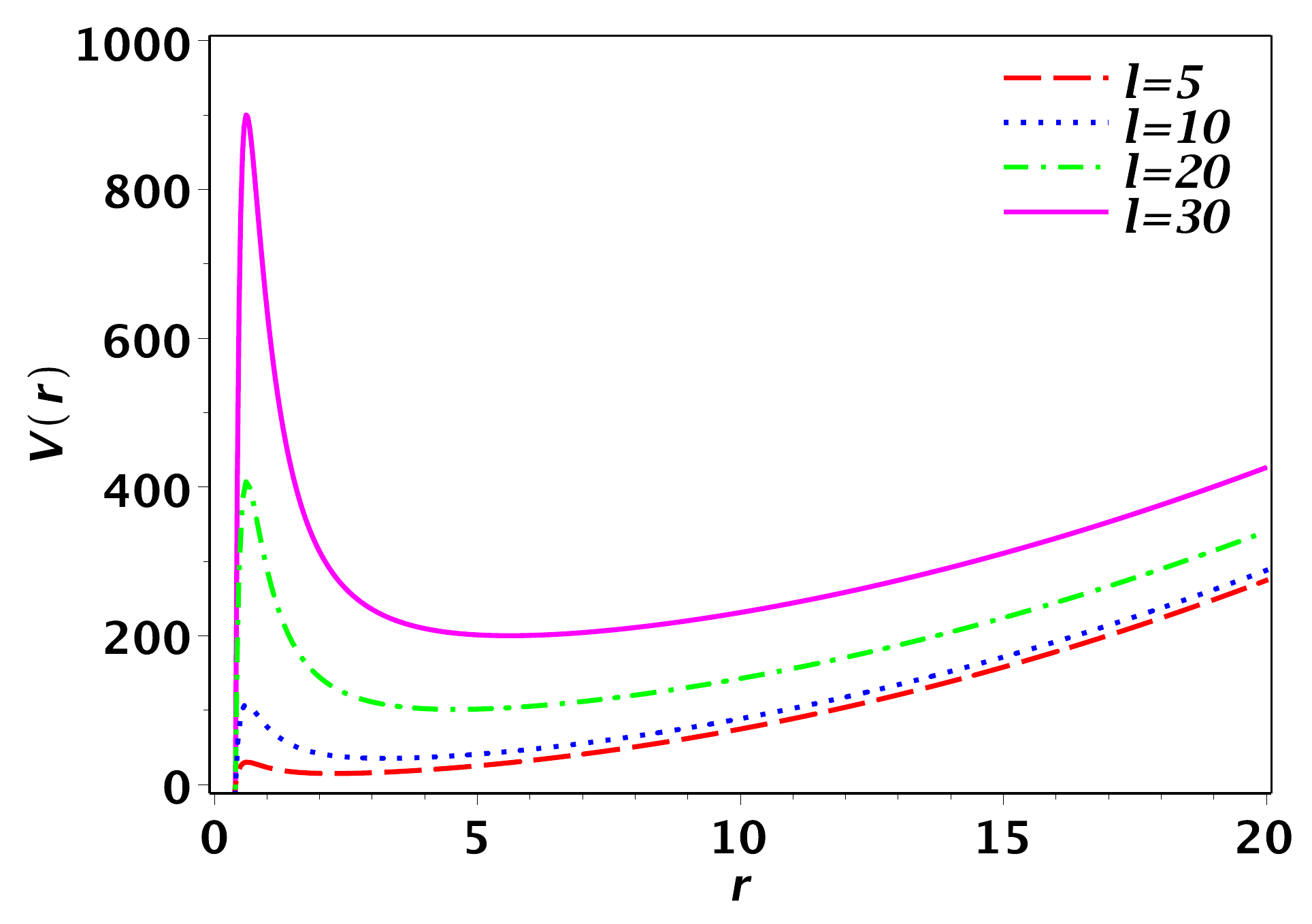}} \nonumber
\end{eqnarray}
\caption{Effective potential as a function of $r$ for axial Proca
  perturbations for $m=2$, fixing $\ell=1$ for several values of $\bar z$
  (left) and fixing $\bar z=0.5$ for several values of $\ell$ (right). The
  event horizon is located at $r_+=0.4$.}
\label{potpro}
\end{figure}
%%%%%%%%%%%%%%%%%%%%%%%%%%%%%%%%%%%%%%%%%%%%%%%%%%%%%%%%%%%%%%%%%%%%%
Figure \ref{potpro} shows the effective potential as a function of $r$
for axial Proca perturbations. From Eq.(\ref{procaeffpotaxial}) it is
easy to see that for large $r$ the mass term dominates, so the
potential diverges as $r^2$, a fact that can also be noticed from the
plots. For large multipole number the potential develops a peak near the event horizon, but it remains always positive definite.

The polar component can be arranged in a set of coupled equations of motion given by
\begin{eqnarray}
\label{procaperturbationpolar}
\frac{F(r)}{\sqrt{F(r)H(r)}}\left[\frac{[j(r,t)_{,t}-f(r,t)_{,r}]r^2}{\sqrt{F(r)H(r)}}\right]_{,r} = [k(r,t)_{,t}-f(r,t)]\ \ell(\ell+1)-m^2r^2 f(r,t)\,, \\
\nonumber\\
\label{procaperturbationpolarI}
\frac{H(r)}{\sqrt{F(r)H(r)}}\left[\frac{[j(r,t)_{,t}-f(r,t)_{,r}]r^2}{\sqrt{F(r)H(r)}}\right]_{,t} = [k(r,t)_{,r}-j(r,t)]\ \ell(\ell+1)-m^2r^2 j(r,t)\,.
\end{eqnarray}
Equations (\ref{procaperturbationpolar}) and
(\ref{procaperturbationpolarI}) cannot be decoupled for arbitrary
values of multipoles $\ell$. However, for the monopole mode $(\ell=0)$,
these equations become decoupled. In fact, this case corresponds to a
scalar mode. Although this is forbidden for the Maxwell field, which has
only two helicities, it is certainly possible for the Proca field (due to
its mass). Thus, rewriting Eqs. (\ref{procaperturbationpolar}) and
(\ref{procaperturbationpolarI}) and performing a  substitution in terms of a new function $N(r,t)$, we have
\begin{eqnarray}
\label{procaperturbationpolarII}
\frac{F(r)}{\sqrt{F(r)H(r)}}[N(r,t)r^2]_{,r} = -m^2r^2 f(r,t)\,, \\
\nonumber\\
\label{procaperturbationpolarIIa}
\frac{H(r)}{\sqrt{F(r)H(r)}}[N(r,t)r^2]_{,t} = -m^2r^2 j(r,t)\,,
\end{eqnarray}
where the function $N(r,t)$ is defined by
\begin{eqnarray}
N(r,t)=\frac{[j(r,t)_{,t}-f(r,t)_{,r}]}{\sqrt{F(r)H(r)}}\,.
\end{eqnarray}
Deriving Eq.(\ref{procaperturbationpolarII}) with respect to $r$ and Eq.(\ref{procaperturbationpolarIIa}) with respect to $t$ and adding the resulting equations we obtain
\begin{eqnarray}
\label{procaperturbationpolarIII}
-\frac{\partial^2 N(r,t)}{\partial t^2}+\frac{\partial^2 N(r,t)}{\partial r_{*}^{2}}+\frac{\partial}{\partial r_*}\left[\frac{2}{r}\sqrt{\frac{F}{H}}N(r,t)\right]-m^2 F N(r,t)=0.
\end{eqnarray}
Setting $N(r,t)=\frac{R(r,t)}{r}$, Eq.(\ref{procaperturbationpolarIII})
turns out to be
\begin{eqnarray}
\label{procaperturbationpolarIV}
-\frac{\partial^2 R(r,t)}{\partial t^2}+\frac{\partial^2 R(r,t)}{\partial r_{*}^{2}}-V_{P}^{polar}(r)R(r,t)=0\,,
\end{eqnarray}
with the effective potential expressed as
\begin{eqnarray}
\label{procapotentialpolar}
V_{P}^{polar}(r)=\left[\frac{2}{r^2}\frac{F}{H}+ m^2 F-\frac{1}{2r}\left(\frac{F}{H}\right)_{,r} \right]\,.
\end{eqnarray}

%%%%%%%%%%%%%%%%%%%%%%%%%%%%%%% FIGURA 3 %%%%%%%%%%%%%%%%%%%%%%%%%%%%%%%%%%%%%%%%%%
\begin{figure}[h]
\begin{eqnarray}
\rotatebox{0}
{\includegraphics[height=.35\textwidth]{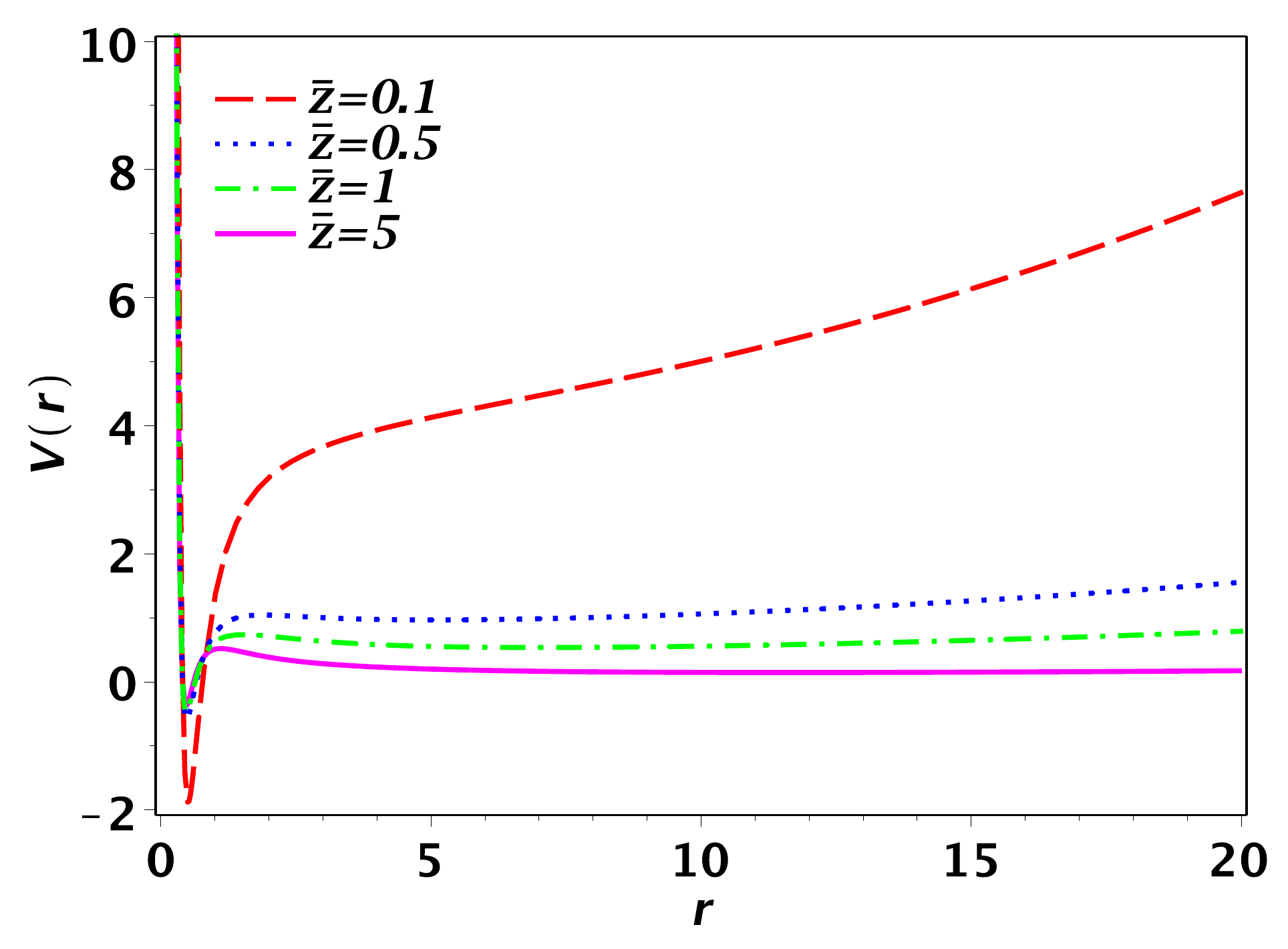}}
\rotatebox{0}
{\includegraphics[height=.35\textwidth]{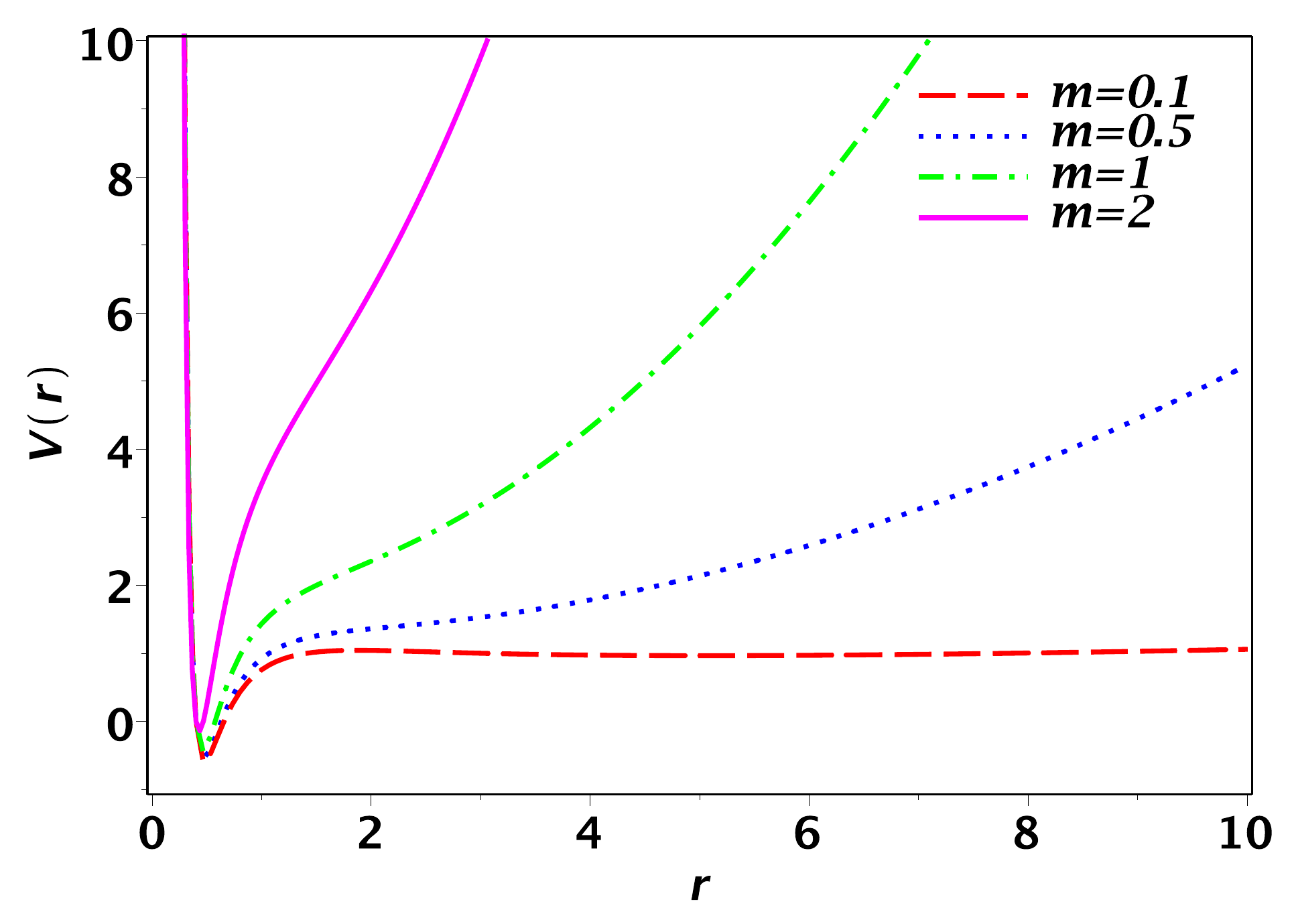}} \nonumber
\end{eqnarray}
\caption{Effective potential as a function of $r$ for polar Proca
  perturbations fixing $m=0.1$ for several values of $\bar z$ (left)
  and fixing $\bar z=0.5$ for several values of Proca field mass $m$ (right). The event horizon is located at $r_+=0.4$.}
\label{potpropolar}
\end{figure}
%%%%%%%%%%%%%%%%%%%%%%%%%%%%%%%%%%%%%%%%%%%%%%%%%%%%%%%%%%%%%%%%%%%%%%%%%%%%%%%%%%%%

Figure \ref{potpropolar} shows the effective potential as a function of
$r$ for polar Proca perturbations. As we can observe, $\bar z$ and $m$
have similar effects on the form of the potential. When one of them is
fixed, and the other parameter modifies the depth and width of the
negative potential well near the event horizon. In fact, the smaller
this varying parameter is, the deeper and wider the well
becomes. Another interesting feature is that there is a range of
values from $0$ to $m_*$ or $\bar z_*$ for which the
well appears right outside the event horizon. When $m>m_*$ or $\bar z
> \bar z_*$, the well is shifted to values $r<r_+$, so it is not
relevant for our study anymore.
Furthermore, as $m$ increases, the potential diverges as $r^2$ in a
faster manner.

A redefinition of the mass in the potentials presented in
Eqs.(\ref{procaeffpotaxial}) and (\ref{procapotentialpolar}) as
$m^2=\mu^2 / \bar z$ makes both, axial and polar massive
electromagnetic perturbations, invariant by the same coordinate
transformation previously discussed.

In the spatial infinity $(r\to\infty)$, axial and polar Proca effective potentials reduce to the same value
\begin{equation}
\label{procapotaxialasymp}
 V_P(r)\sim \frac{m^2 r^2}{12 \bar{z}}\,,
\end{equation}
which is consistent with the graph analysis. Also, different from the
Maxwell case, both components of the Proca field $\Psi$ will become
\begin{equation}
\label{procaaxialasymp}
\Psi(r)=\frac{C_1}{r^{\alpha_{+}}}+\frac{C_2}{r^{\alpha_{-}}}\,,
\end{equation}
where
\begin{equation}
\alpha_{\pm}=\frac{1}{2}\pm\frac{\sqrt{1+9m^2\bar{z}}}{2}\,.
\end{equation}
In this case we also obtain the above-mentioned symmetry for Maxwell
perturbations by rescaling the spacetime variables $t=\tau\sqrt{\bar z}$ and
$r=x\sqrt{\bar z}$ and the black hole mass as $M=M^\prime \sqrt{\bar
  z}$, such that (for a given Proca mass) the quasinormal mode
frequency scales as
\begin{equation}\label{rescale}
\omega =\frac1{\sqrt z} \, {\bf f}\left(\frac{r_+}{\sqrt z}\right)\,,
\end{equation}
where ${\bf f}$ is a real function. Such a result can explicitly be checked
in the corresponding tables shown in the Appendix.

%%%%%%%%%%%%%%%%%%%%%%%%%%%%%%%%%%%%%%%%%%%%%%%%%%%
\section{Fermionic field perturbation}  \label{fermionicperturbation}
%%%%%%%%%%%%%%%%%%%%%%%%%%%%%%%%%%%%%%%%%%%%%%%%%

One of the most interesting possibilities is the introduction of
fermions in the model. Let us consider a spinorial field $\Psi$ with mass $\mu_s$ as a perturbation in the spacetime given by Eq.(\ref{metric}), obeying Dirac equation,
\begin{equation}\label{dirac}
i \gamma^{(a)} e_{(a)} ^\mu \nabla_\mu \Psi - \mu_s \Psi = 0 \,,
\end{equation}
where the covariant derivative is defined according to
\begin{equation}\label{covariant}
\nabla_\mu = \partial_\mu + \frac{1}{4} \omega_\mu ^{(a)(b)} \gamma_{[a} \gamma_{b]}\,,
\end{equation}
and $\omega_\mu ^{(a)(b)}$ is the spin connection written in terms of the tetrad basis $e_\mu ^{(a)}$ as
\begin{equation}\label{spincon}
\omega_\mu ^{(a)(b)} = e_\nu ^{(a)} \partial_\mu e^{(b)\nu} + e_\nu ^{(a)} \Gamma^\nu _{\mu\sigma} e^{\sigma(b)} \,.
\end{equation}
Here greek indices refer to spacetime coordinates $(t,r,\theta,\phi)$, while latin indices enclosed in parentheses are assigned to flat tangent space where the tetrad basis is defined. The Dirac matrices $\gamma^{(a)}$ are given by
\begin{equation}\label{dirmat}
\gamma^0 = \left( \begin{array}{cc}
           -i & 0 \\
           0  & i \end{array} \right) \,, \quad
\gamma^i=  \left( \begin{array}{cc}
           0 & -i\sigma^i \\
           i\sigma^i & 0 \end{array} \right)\,, \quad
i=1,2,3\,,
\end{equation}
where $\sigma^i$ are the Pauli matrices. Using metric (\ref{metric}) we can specify the tetrad basis as
\begin{equation}
e_0 ^{(a)} = \sqrt{F}\, \delta_{\phantom{..}0}^{(a)}\,, \quad e_1 ^{(a)} = \sqrt{G}\, \delta_{\phantom{..}1}^{(a)}\,, \quad e_2 ^{(a)} = r \, \delta_{\phantom{..}2}^{(a)}\,, \quad e_3 ^{(a)} = r\,\sin\theta \, \delta_{\phantom{..}3}^{(a)}\,.
\end{equation}

Thus, the Dirac equation (\ref{dirac}) becomes
\begin{equation}\label{dir1}\nonumber
\left[ \gamma^0 \left( \frac{1}{\sqrt{F}}\right)  \partial_t + \gamma^1 \left(\frac{1}{\sqrt{H}}\right) \left( \partial_r + \frac{1}{4}\frac{F'}{F} + \frac{1}{r}\right) + \gamma^2 \left(\frac{1}{r}\right) \left( \partial_\theta + \frac{1}{2} \cot\theta \right)
+ \gamma^3 \left(\frac{1}{r\sin\theta}\right) \partial_\phi + i\mu_s \right] \Psi = 0 \,.
\end{equation}

In order to simplify this equation, let us define
\begin{equation}\label{subs1}
\Psi(t,r,\theta,\phi) = F(r)^{-1/4} \Phi(t,r,\theta,\phi) \,;
\end{equation}
thus, Eq.(\ref{dir1}) becomes
\begin{equation}\label{dir2}
\left[ \gamma^0 \left(\frac{1}{\sqrt{F}}\right) \partial_t + \gamma^1 \left(\frac{1}{\sqrt{H}}\right) \left(\partial_r + \frac{1}{r} \right) + \gamma^2 \left(\frac{1}{r}\right) \left(\partial_\theta + \frac{1}{2} \cot\theta\right) + \gamma^3 \left(\frac{1}{r\sin\theta} \right) \partial_\phi + i\mu_s \right] \Phi = 0\,.
\end{equation}
Notice that when $F(r)=H(r)^{-1}=(1-2M/r)$ Eq.(\ref{dir2}) reduces to the Schwarzschild case~\cite{cho}.

By decomposing Dirac equation in an angular and a two-dimensional ($t,r$) part~\cite{lopez}, for a two-spinor the latter equation reads
\begin{eqnarray}
\left(\partial_t - \sqrt{\frac{F}{H}} \partial_r\right) \psi_2 &=& \left( i\kappa \frac{\sqrt{F}}{r} + i\mu_s \sqrt{F}\right) \psi_1 \label{2spina}\\
\left(\partial_t + \sqrt{\frac{F}{H}} \partial_r \right)\psi_1 &=& -\left( i\kappa \frac{\sqrt{F}}{r} + i\mu_s \sqrt{F}\right) \psi_2 \label{2spinb}\,,
\end{eqnarray}
where $\kappa$ is a constant associated to the variable separation that can be expressed as $\kappa = i(\ell +1) \equiv iK$. By writing the two-spinor components as
\begin{eqnarray}
\psi_1 &=& e^{-i\omega t} R_1(r)\\
\psi_2 &=& e^{-i\omega t} R_2(r)\,,
\end{eqnarray}
and switching to the tortoise coordinate (\ref{tartaruga1}), Eqs.(\ref{2spina}) and (\ref{2spinb}) become
\begin{eqnarray}
\left( \frac{d}{dr_*} +i\omega \right) R_2 &=& \sqrt{F} \left(\frac{K}{r} +i\mu_s\right) R_1 \label{2spin2a}\\
\left( \frac{d}{dr_*} -i\omega \right) R_1 &=& \sqrt{F} \left(\frac{K}{r} -i\mu_s\right) R_2\,. \label{2spin2b}
\end{eqnarray}
Now, we define a new function $\theta$, set a new tortoise coordinate $\hat r_*$, and rescale again the spinorial components $R_1$ and $R_2$ through the expressions
\begin{eqnarray}
\theta = \arctan\left(\frac{\mu_s r}{K} \right) \,, &\qquad& \hat r_* = r_* + \frac{1}{2\omega} \arctan\left(\frac{\mu_s r}{K} \right)\,, \label{tartaruga2}\\
R_1 = e^{-i\theta/2}\Phi_1\, {\hbox{and}} &\qquad& R_2 = e^{i\theta/2}\Phi_2\,.
\end{eqnarray}
Thus, Eqs.(\ref{2spin2a}) and (\ref{2spin2b}) turn out to be
\begin{equation}\label{2spin3}
\left(\frac{d}{d\hat r_*} \pm i\omega \right) \Phi_{\binom{2}{1}} = W \Phi_{\binom{1}{2}} \,,
\end{equation}
where the so-called superpotential can be written as
\begin{equation}\label{superpot}
W = \frac{\left[F\left(K^2/r^2 +\mu_s^2\right)\right]^{1/2}}{1+\frac{\mu_s K}{2\omega(K^2+\mu_s^2r^2)}\sqrt{\frac{F}{H}}}\,.
\end{equation}
Finally, in order to express our result in a more familiar way in terms of the superpartner potentials, let us define
\begin{equation}
Z_\pm = \Phi_1 \pm \Phi_2 \, .
\end{equation}
Thus, Eqs.(\ref{2spin3}) can be brought to their final form,
\begin{equation}
\left( \frac{d^2}{d\hat r_*^2} + \omega^2 \right) Z_\pm = V_\pm Z_\pm \,,
\end{equation}
with the superpartner potentials given by
\begin{equation}\label{superpartner}
V_\pm = W^2 \pm \frac{dW}{d\hat r_*} \,.
\end{equation}

In what follows, we will consider the case of a massless fermion field.
In this case the superpotential (\ref{superpot}) reduces to
\begin{equation}
W=\sqrt{F}\frac{K}{r}\,.
\end{equation}
The superpartner potentials $V_+$ and $V_-$ yield the same quasinormal spectrum since they satisfy the relation
\begin{equation}
V_+ -V_- -2\frac{dW}{dr_*} = 0 \,.
\end{equation}
Notice that both tortoise coordinates coincide when $\mu_s=0$.

%%%%%%%%%%%%%%%%%%%%%%%%%%%%%%%%%%%%%% FIGURA 4 %%%%%%%%%%%%%%%%%%%%%%%%%%%%%%
\begin{figure}[h!]
\begin{eqnarray}
\rotatebox{0}
{\includegraphics[width=.45\textwidth]{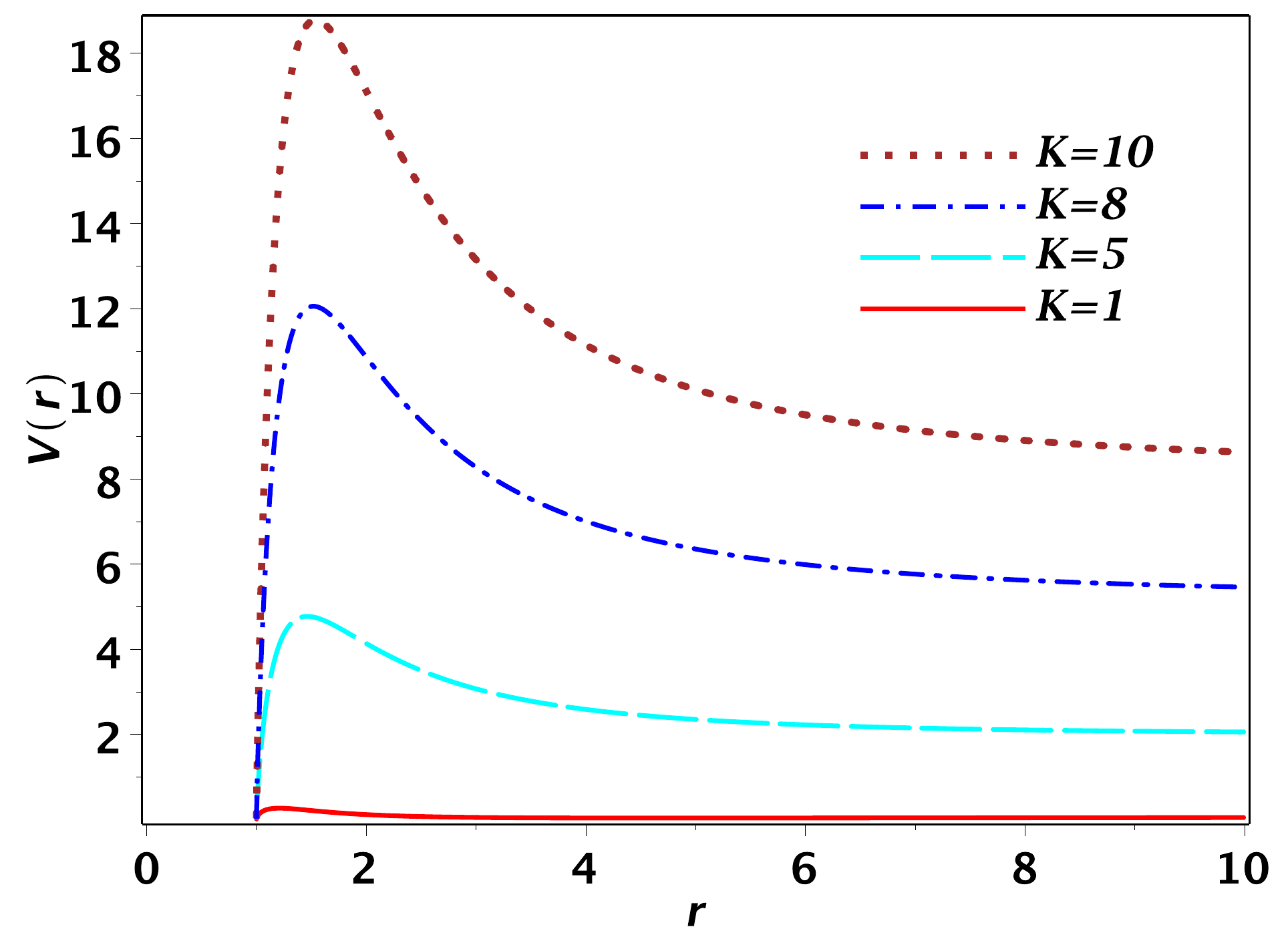}} \quad
{\includegraphics[width=.45\textwidth]{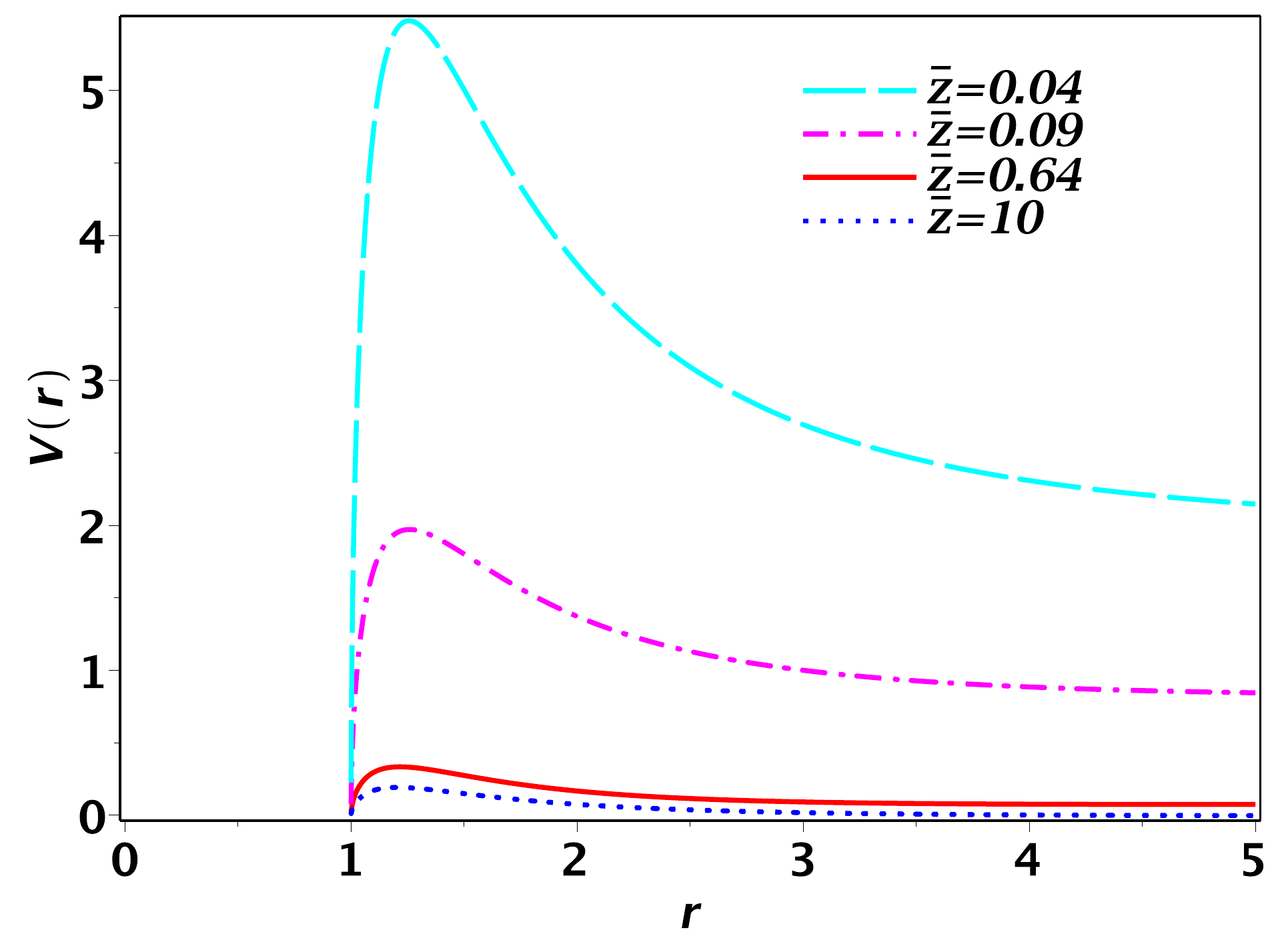}}\nonumber
\end{eqnarray}
\caption{Effective potential $V_+$ as a function of $r$ for spinorial
  perturbations fixing $\bar z=1$ for different $K$ (left) and fixing $K=1$ for different $\bar z$ (right). Notice that event the horizon is located at $r_+=1$.}
\label{potspin+}
\end{figure}

%%%%%%%%%%%%%%%%%%%%%%%%% FIGURA 5 %%%%%%%%%%%%%%%%%%%%%%%%%%%%%%%%%%%%%%%%%%
\begin{figure}[h!]
\begin{eqnarray}
\rotatebox{0}
{\includegraphics[width=.45\textwidth]{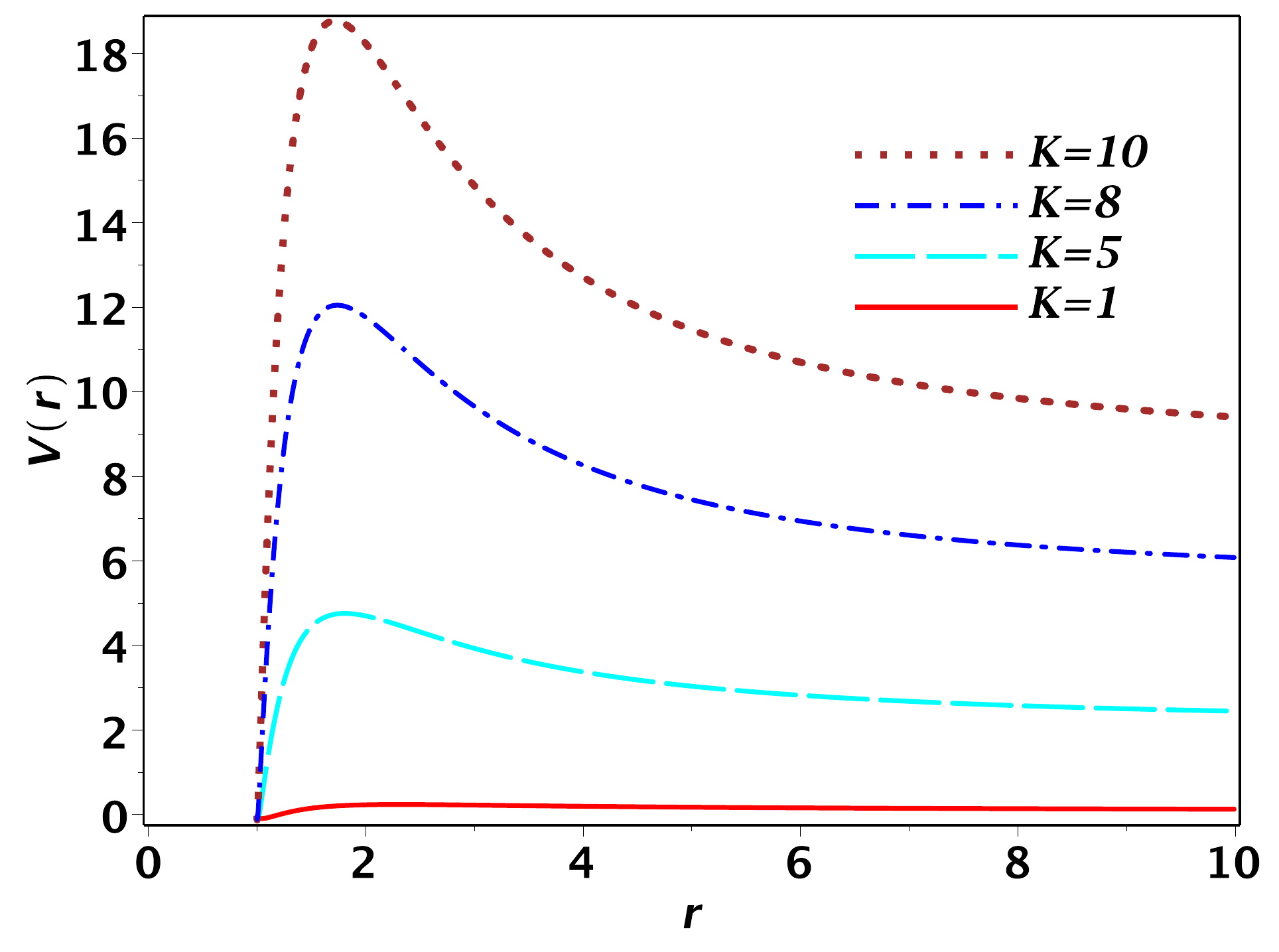}} \quad
{\includegraphics[width=.45\textwidth]{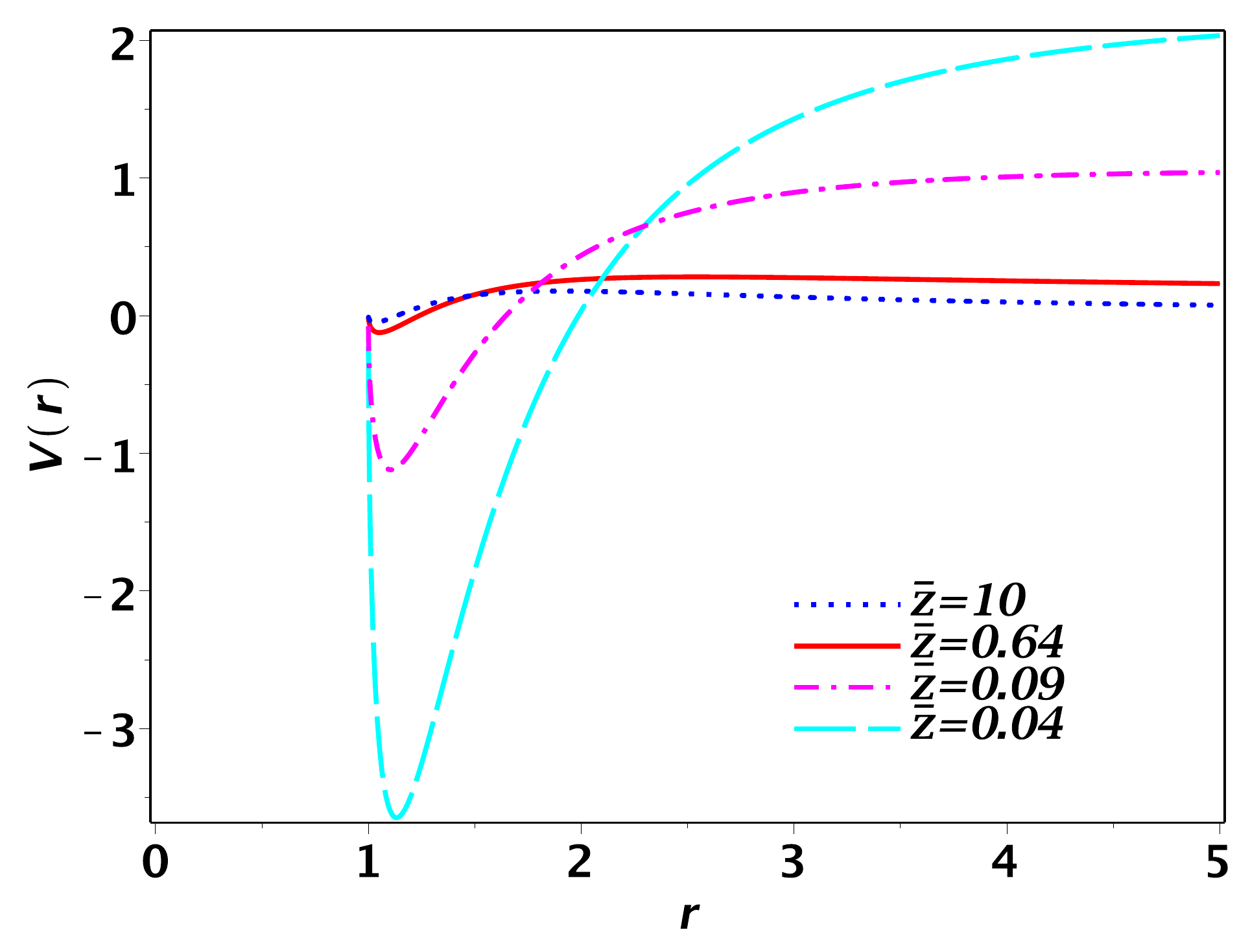}}\nonumber
\end{eqnarray}
\caption{Effective potential $V_-$ as a function of $r$ for spinorial
  perturbations fixing $\bar z=1$ for different $K$ (left) and fixing $K=1$ for different $\bar z$ (right). Notice that the event horizon is located at $r_+=1$.}
\label{potspin-}
\end{figure}
%%%%%%%%%%%%%%%%%%%%%%%%%%%%%%%%%%%%%%%%%%%%%%%%%%%%%%%%%%%%%%%%%%%%%%%%%
From Figs. \ref{potspin+} and \ref{potspin-}, we see that superpartner
effective potentials display a maximum (or minimum in the case of
$V_-$ for small $\bar z$) around the event horizon neighborhood and decrease to a constant value as the radial coordinate goes to infinity,
\begin{equation}
\lim_{r\rightarrow \infty}V_{+} = \lim_{r\rightarrow \infty}V_{-}
\rightarrow \frac{K^{2}}{12 \bar z}\,.
\end{equation}

For both potentials, as $K$ grows, a peak rises up. On the other
hand, as the $\bar z$ parameter increases, the peak in $V_+$ or the well
in $V_-$ decreases and
gradually reaches the curves corresponding to nonmassive
fermions propagating in a Schwarzschild solution. This result
perfectly agrees with the fact that metric (\ref{metric}) approaches
the Schwarzschild solution in the limit $\bar z \rightarrow\infty$.

From this behavior it is clear that we can apply a WKB method to
obtain quasinormal frequencies. It is well known that the WKB method has a
perfect convergence when the parameter associated to the angular
momentum is large compared to the overtone number. In other cases we
must analyze  other parameters to reach some conclusion.

Looking at Fig. \ref{potspin-}, it is clear that the potential $V_-$ is a
very curious case. As $\bar z$ becomes smaller, a negative well develops
and some instabilities could be expected. However, no instability was
found in our numerical calculation. In order to explain this fact, we
approximated the region near the $V_-$ minimum as a harmonic
oscillator potential and found the ground state energy of the
associated state. Performing this procedure numerically, we discovered that this
energy is always larger than the depth of the well. Thus, we have no
bound states, and no unstable mode can exist.

%%%%%%%%%%%%%%%%%%%%%%%%%%%%%%%%%%%%%%%%%%%%%%%%%
\section{Numerical Results} \label{numex}
%%%%%%%%%%%%%%%%%%%%%%%%%%%%%%%%%%%%%%%%%%%%%%%%%

\subsection{Massless vector field}

Let us begin our discussion by showing the results for Maxwell perturbations.

\begin{figure}[h!]
\begin{eqnarray}
\rotatebox{0}
{\includegraphics[width=.5\textwidth]{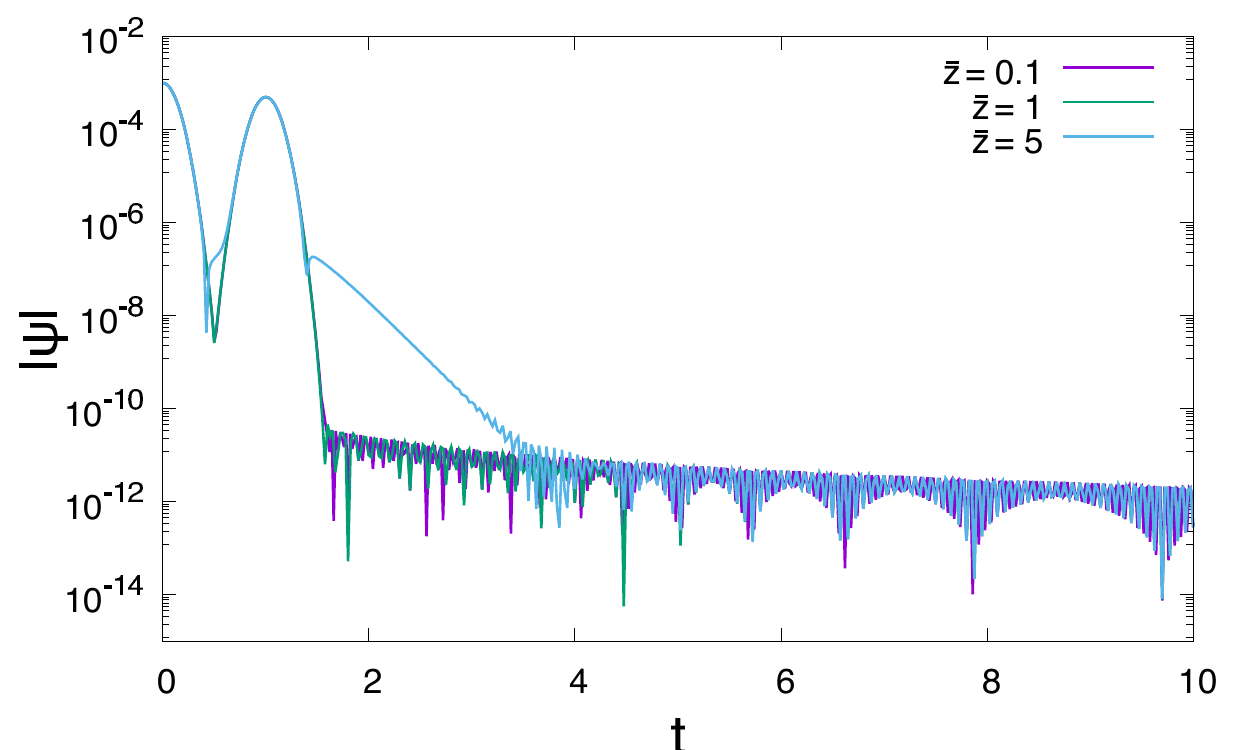}}
\rotatebox{0}
{\includegraphics[width=.5\textwidth]{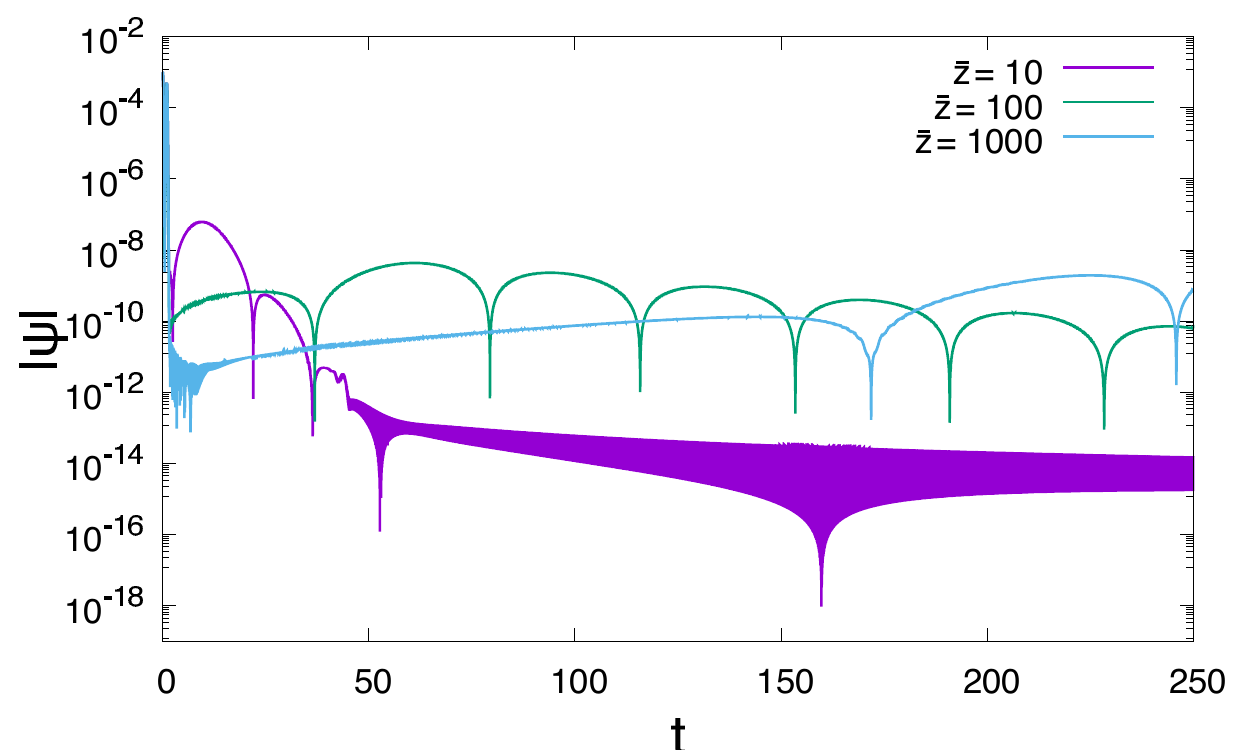}}
\nonumber
\end{eqnarray}
\caption{Maxwell perturbations for different values of $\bar z$ with
  multipole number $\ell=1$ and the event horizon fixed at $r_+=100$
  (left) and $r_+=10$ (right).}
\label{pertmax}
\end{figure}

In Fig.\ref{pertmax}, we show some of our results for small and big values of
$\bar z$ compared to the event horizon with multipole number
$\ell=1$. From these figures we can see that for small values of $\bar z$ modes appear to be stable and display oscillating tails. Although these
perturbations are massless, these tails are a strong indication that
$\bar z$ plays the role of mass for the perturbation. We can also
notice that when $\bar z$ gets bigger, the imaginary part of the
frequencies decreases but it does not reach zero. We also verified
that for bigger multipole numbers $\ell$ modes with $\bar z$ of the
order of the event horizon produce beats and the oscillating tails
decay more slowly. Thus, we can conclude that the model is
stable under Maxwell perturbations.

We also used the Horowitz-Hubeny (HH) method~\cite{hh} in order to numerically obtain the
quasinormal frequencies. Our results are shown in Tables \ref{tabv1}
and \ref{tabv2} in the Appendix.
For small black holes $(r_+ < 1)$ when $\bar z>r_+$ the task of finding
quasinormal frequencies becomes harder and the convergence of the HH
method becomes worse.
We can see that a critical value of $\bar z_{cI}$, indicated with a
$*$ in the tables, satisfies the following
relation $\bar z_{cI} \sim 0.0178\ r_{+}^{2}$. For $\bar z< \bar
z_{cI}$ quasinormal modes become purely imaginary like in a damped
harmonic oscillator.

\subsection{Massive vector field}

Now let us turn our attention to Proca perturbations.
\begin{figure}[t]
\begin{eqnarray}
\rotatebox{0}
{\includegraphics[width=.5\textwidth]{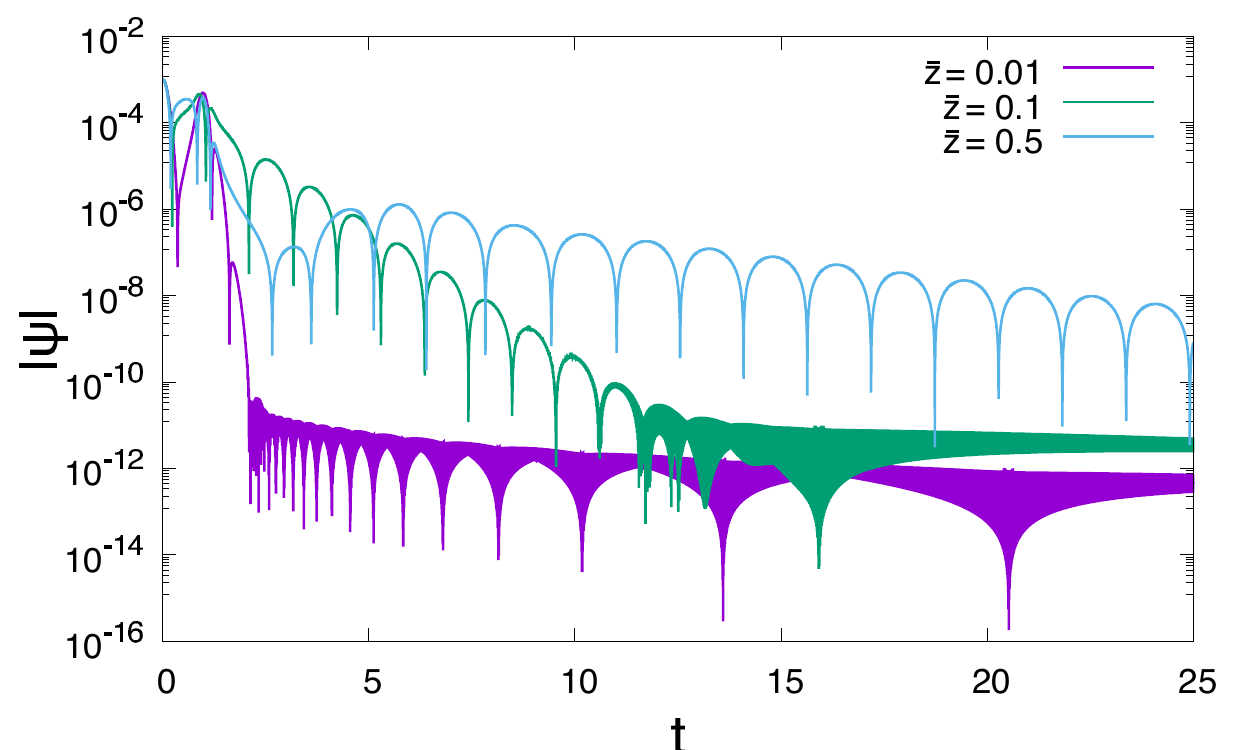}}
\rotatebox{0}
{\includegraphics[width=.5\textwidth]{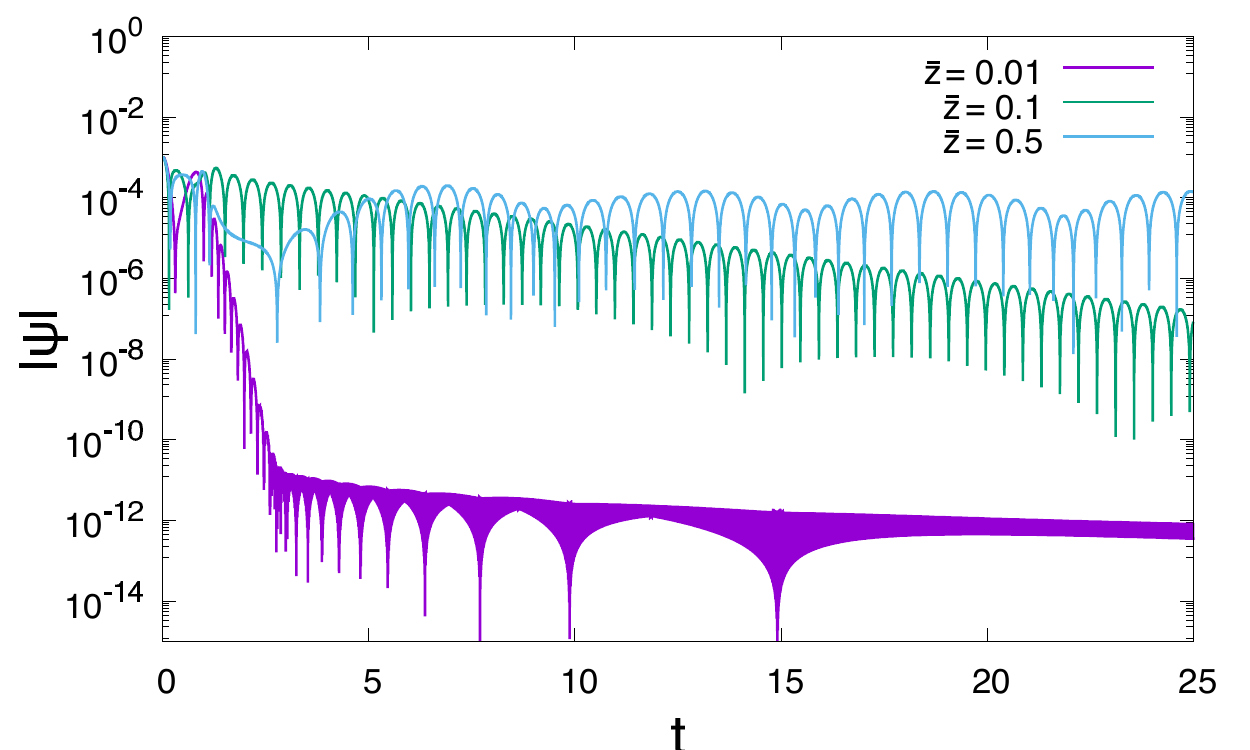}} \nonumber
\end{eqnarray}
\caption{Axial Proca perturbations for $m=2$ when $\ell=1$ (left) and $\ell=5$
  (right) for different values of $\bar z$. In both cases we set the
  event horizon at $r_+=0.4$.}
\label{pertpro}
\end{figure}
In Fig.\ref{pertpro}, we show our results for axial Proca perturbations
fixing the vector field mass $m=2$ for several values of $\bar z$.

Our results establish that modes with small multipole number $\ell$
are always decaying pointing out the
stability of the model under this kind of perturbation. For given
values of $r_+$, $m$ and $\ell$, as $\bar z$ grows, the modes decay in
a slower manner and some of them present oscillating beats and
oscillating tails. Also, as we can infer from Eq.(\ref{procaeffpotaxial}),
if $m$ increases, the modes are damped more rapidly as the potential
is dominated by the $m$ term.

When we consider larger multipole numbers $\ell$ and $\bar z$ around
the same order of the event horizon, long-living nondamped
oscillating modes, the so-called quasiresonant modes
(QRM)~\cite{jap,Konoplya3,Chang}, begin to appear. This can be
understood by looking at the corresponding potential shape in
Fig.\ref{potpro}. As $\ell$ grows, a positive well appears in the
potential making possible the appearance of modes which are trapped
inside the well and begin to oscillate with real frequency.

Table \ref{tabv3} in the Appendix shows the quasinormal frequencies obtained by the HH
method. As mentioned in previous sections, we can clearly see a
symmetry by rescaling the spacetime variables as well as the black
hole mass, such that the frequencies fulfill Eq.(\ref{rescale}).

\begin{figure}[h!]
\begin{eqnarray}
\rotatebox{0}
{\includegraphics[width=.5\textwidth]{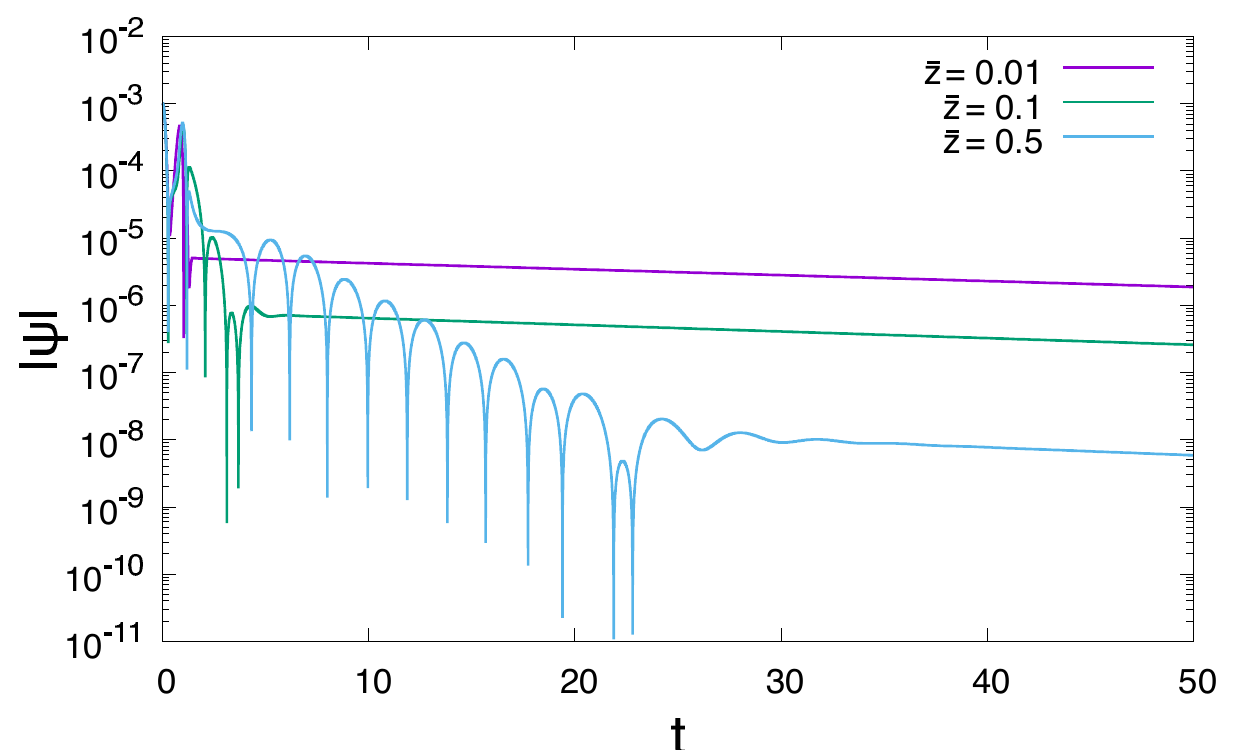}}
\rotatebox{0}
{\includegraphics[width=.5\textwidth]{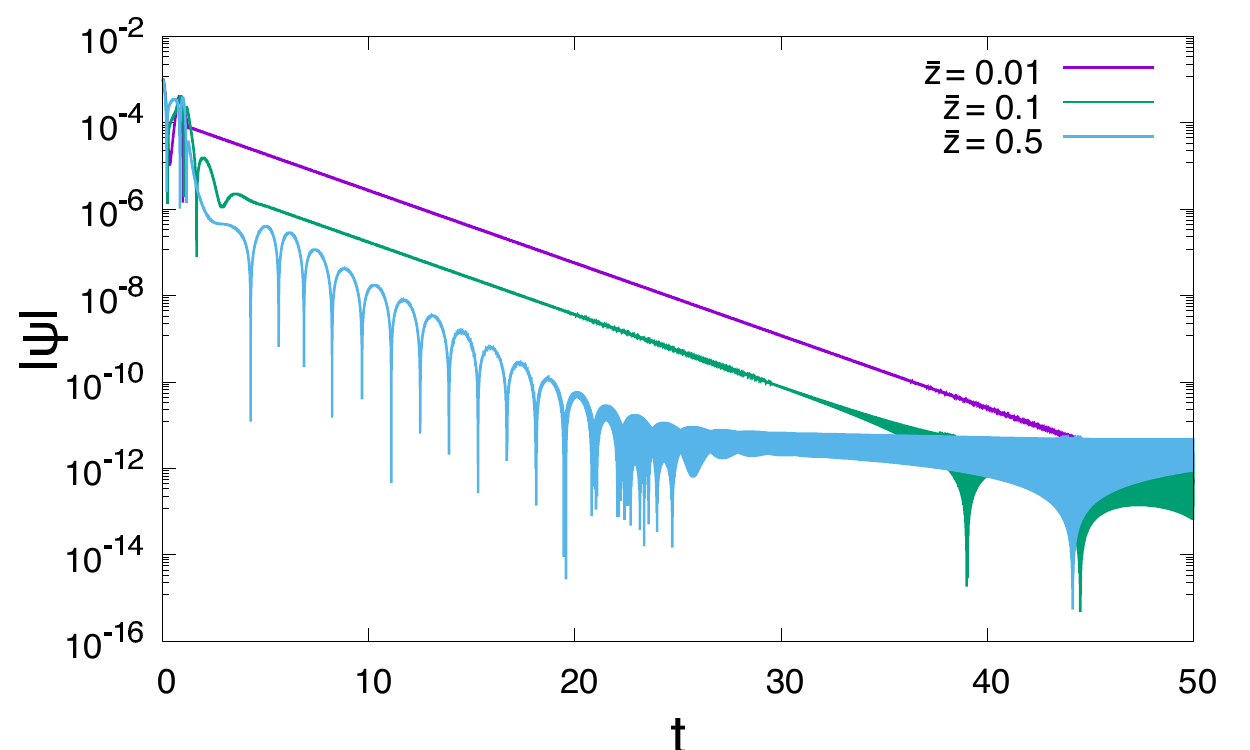}}\nonumber
\end{eqnarray}
\caption{Polar Proca perturbations for $m=0.1$ (left) and
$m=2$ (right) for different values of $\bar z$. In both
  cases, we set the event horizon at $r_+=0.4$.}
\label{pertpol}
\end{figure}

\bigskip

In the case of polar Proca modes we can see the evolution of the
perturbations in Fig.\ref{pertpol}. We can observe that for small
field mass $m$ perturbations decay more slowly than for bigger mass. In 
addition, for small values of
$\bar z$ compared to the event horizon we can check that after a fast
decay there is a power-law tail. And when $\bar z$ gets of the same
order of the event horizon, oscillating beats appear. Thus, the model
is stable under this kind of perturbation.

In Table \ref{tabv4} in the Appendix our results using the HH method are displayed. Again,
there is a critical value $\bar z_{cI}$ below which the modes become
purely imaginary.

%%%%%%%%%%%%%%%%%%%%%%%%%%%%%%%%%%%%%%%%%%%%%%%%%%%

\subsection{Massless spinorial field and Klein paradox}

\begin{figure}[h!]
\begin{eqnarray}
\rotatebox{0}
{\includegraphics[width=.5\textwidth]{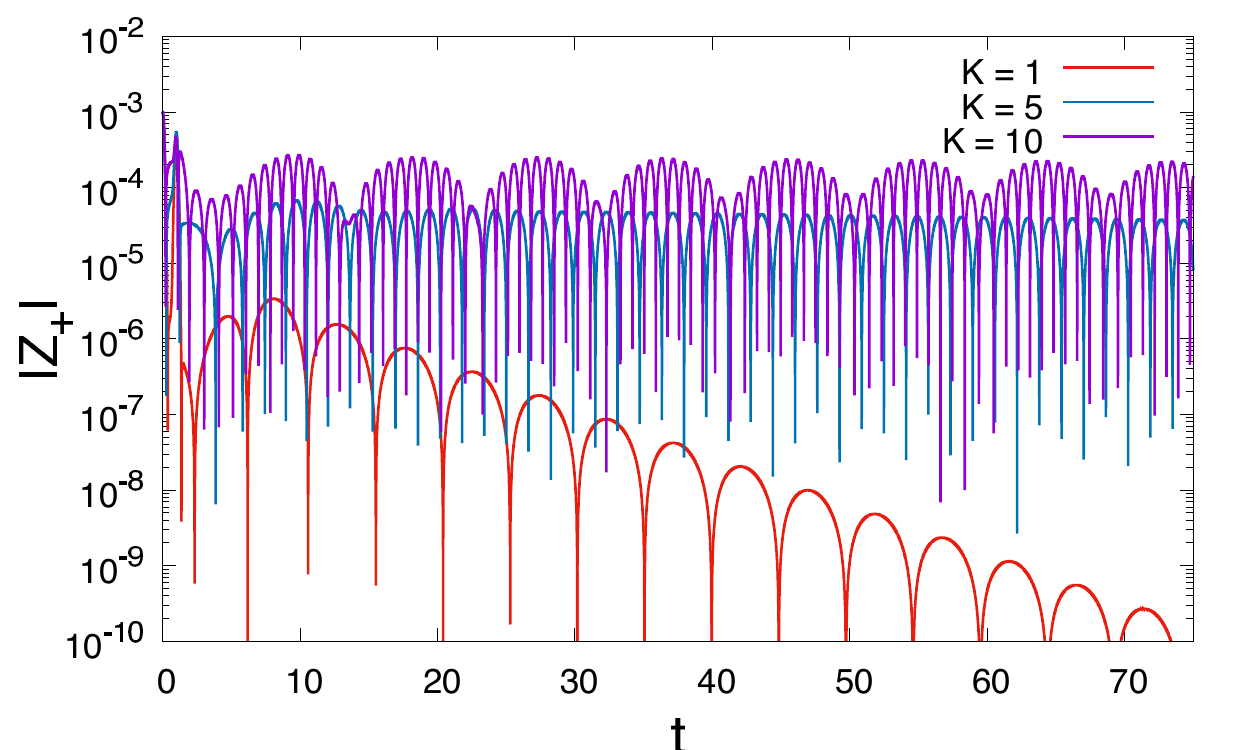}}
\rotatebox{0}
{\includegraphics[width=.5\textwidth]{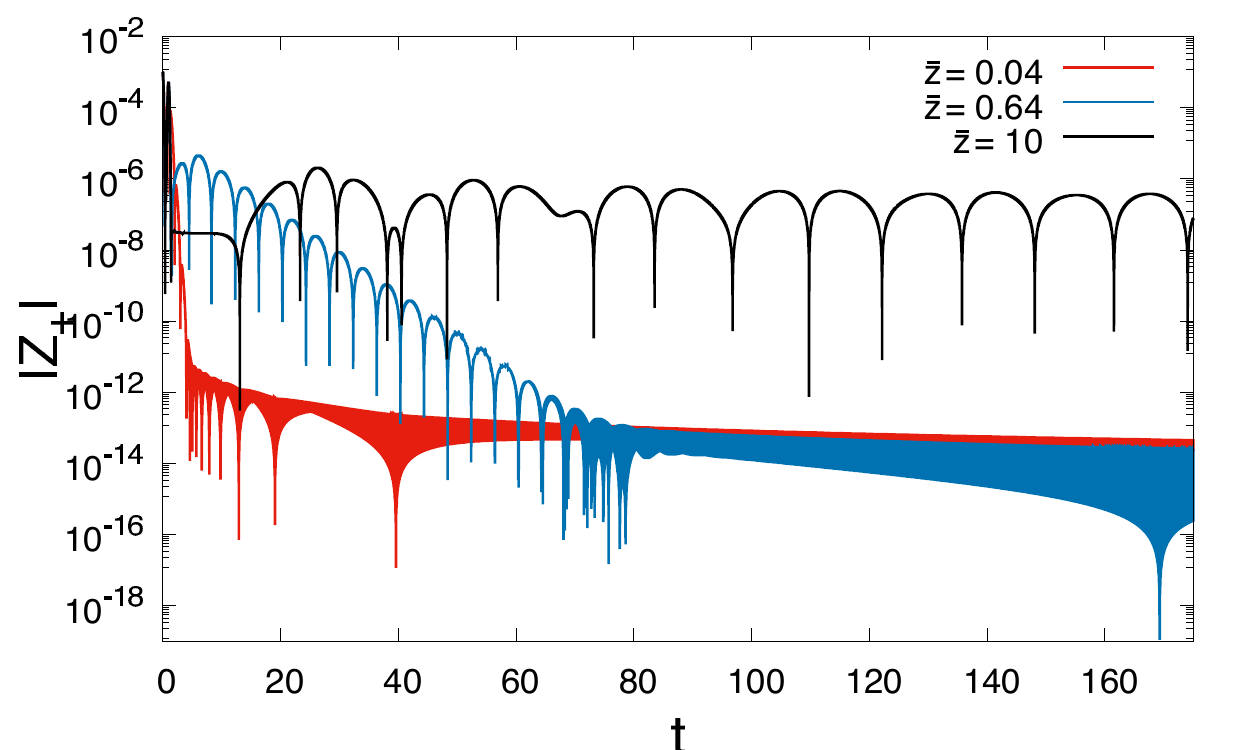}}\nonumber
\end{eqnarray}
\caption{Spinorial perturbations for superpartner potential $V_+$ for
  different values of $K$ fixing $\bar z=1$ (left) and for different
  values of $\bar z$ fixing $K=1$ (right). In both
  cases we set the event horizon at $r_+=1$.}
\label{spgraphP}
\end{figure}

\begin{figure}[h!]
\begin{eqnarray}
\rotatebox{0}
{\includegraphics[width=.5\textwidth]{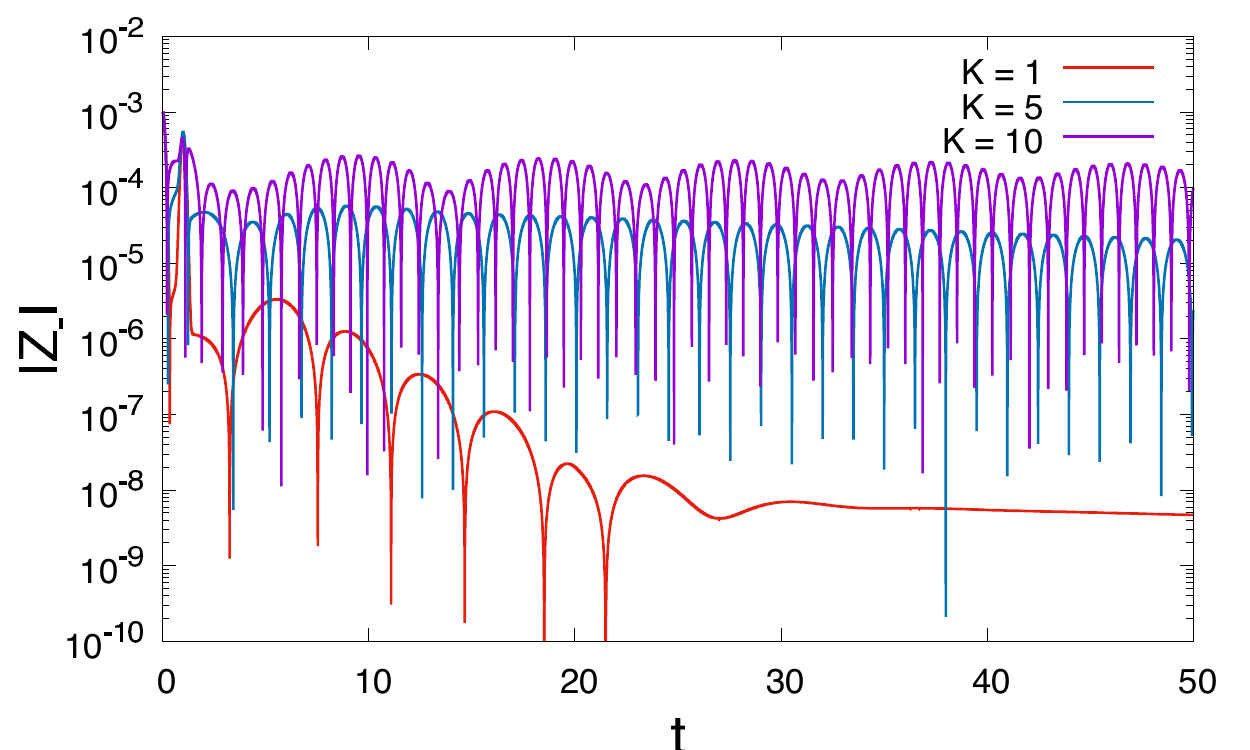}}
\rotatebox{0}
{\includegraphics[width=.5\textwidth]{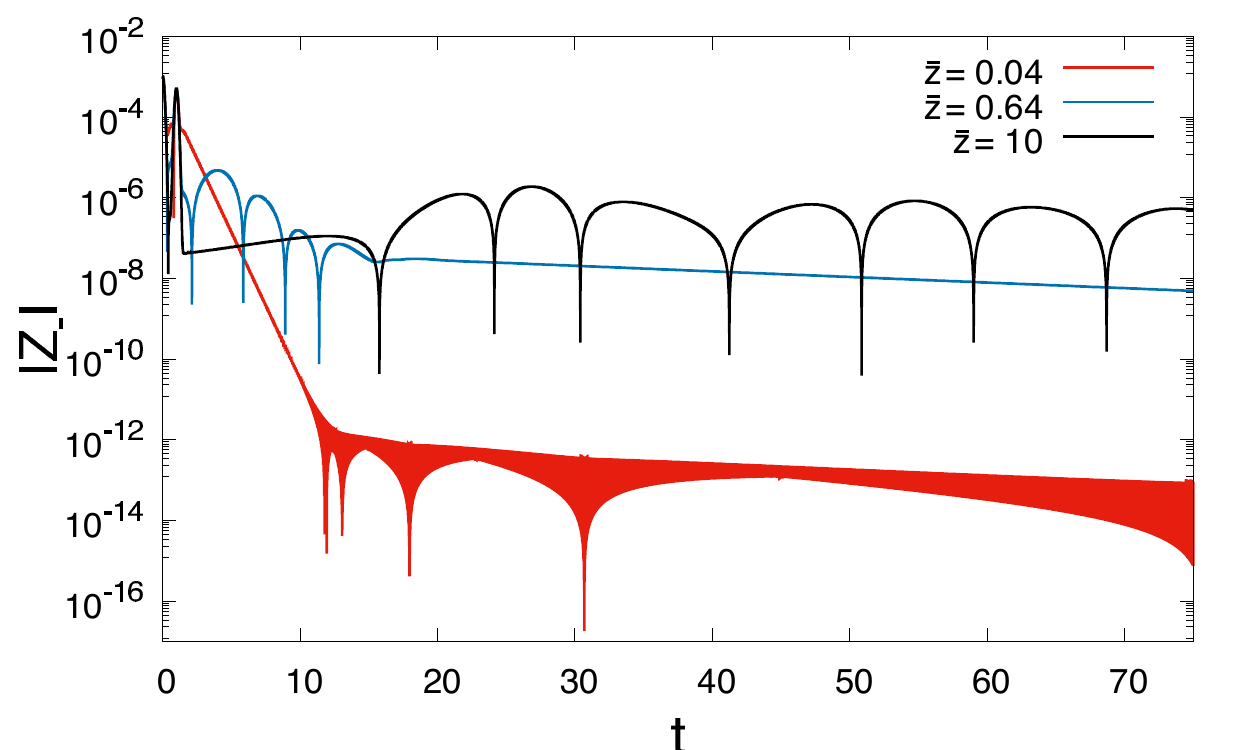}}\nonumber
\end{eqnarray}
\caption{Spinorial perturbations for superpartner potential $V_-$ for
  different values of $K$ fixing $\bar z=1$ (left) and for different
  values of $\bar z$ fixing $K=1$ (right). In both
  cases we set the event horizon at $r_+=1$.}
\label{spgraphM}
\end{figure}

Our results for quasinormal modes are displayed in Fig.\ref{spgraphP}
for the superpartner potential $V_+$ and Fig.\ref{spgraphM} for the
superpartner potential $V_-$. They show that
the model is stable for both $\bar z/r_+\lesssim 1$ and large $\bar
z$. In particular, notice that we did not find any unstable mode for the
Schwarzschild case, which agrees with the result shown in Ref. \cite{cho}. As $\bar{z}$ grows, the imaginary part of the frequency gets smaller, but is still negative, so the modes decay more slowly. Moreover, when $\bar{z}\rightarrow\infty$, i.e., in the Schwarzschild limit, perturbations always decay with an oscillating tail. Both effects the longer-living modes and the oscillating tails have been related to the mass of the perturbation in other models (see Refs. \cite{knoll1,knoll,konoplya-zhidenko} and references therein). Although we are dealing here with massless spinors, the responsible for both behaviors is the $\bar{z}$ term that behaves like a mass term in the Lagrangian. Thus, our results perfectly agree with the well-known behavior in Schwarzschild spacetime, in which massive perturbations have oscillating tails. 

\begin{figure}[h!]
\begin{eqnarray}
\rotatebox{0}
{\includegraphics[width=.49\textwidth]{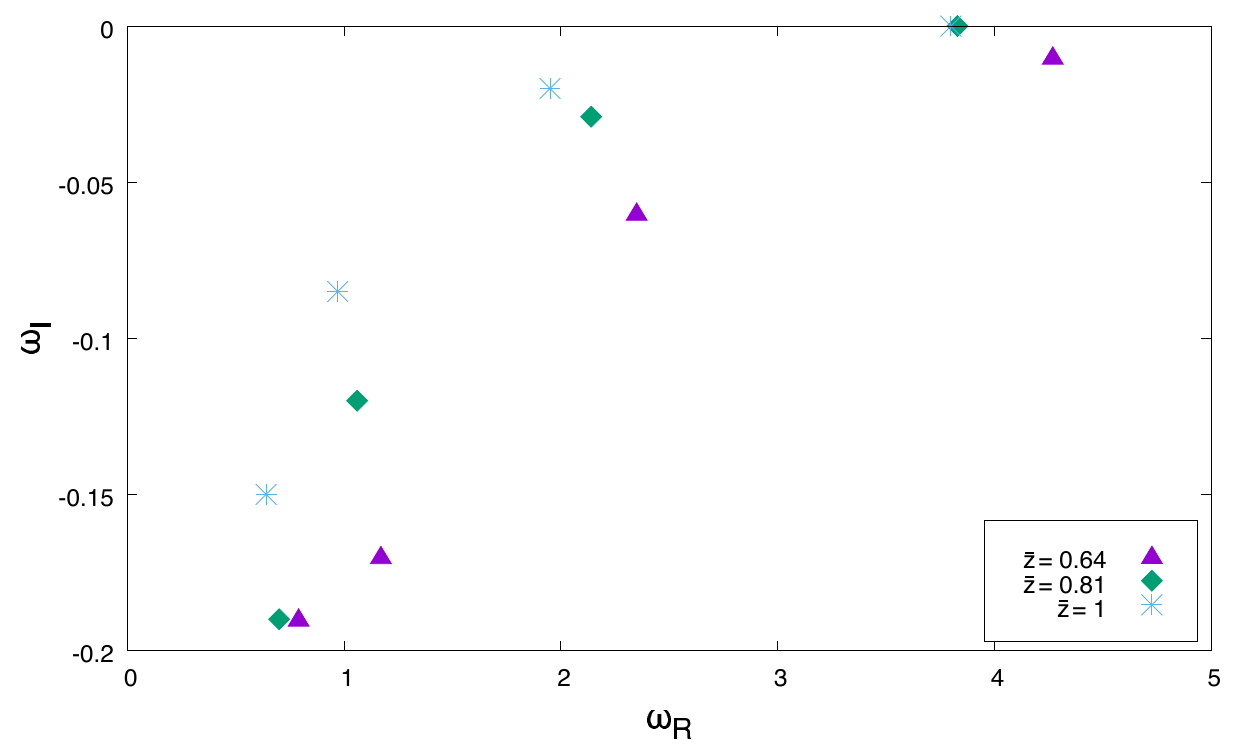}}\quad
\rotatebox{0}
{\includegraphics[width=.49\textwidth]{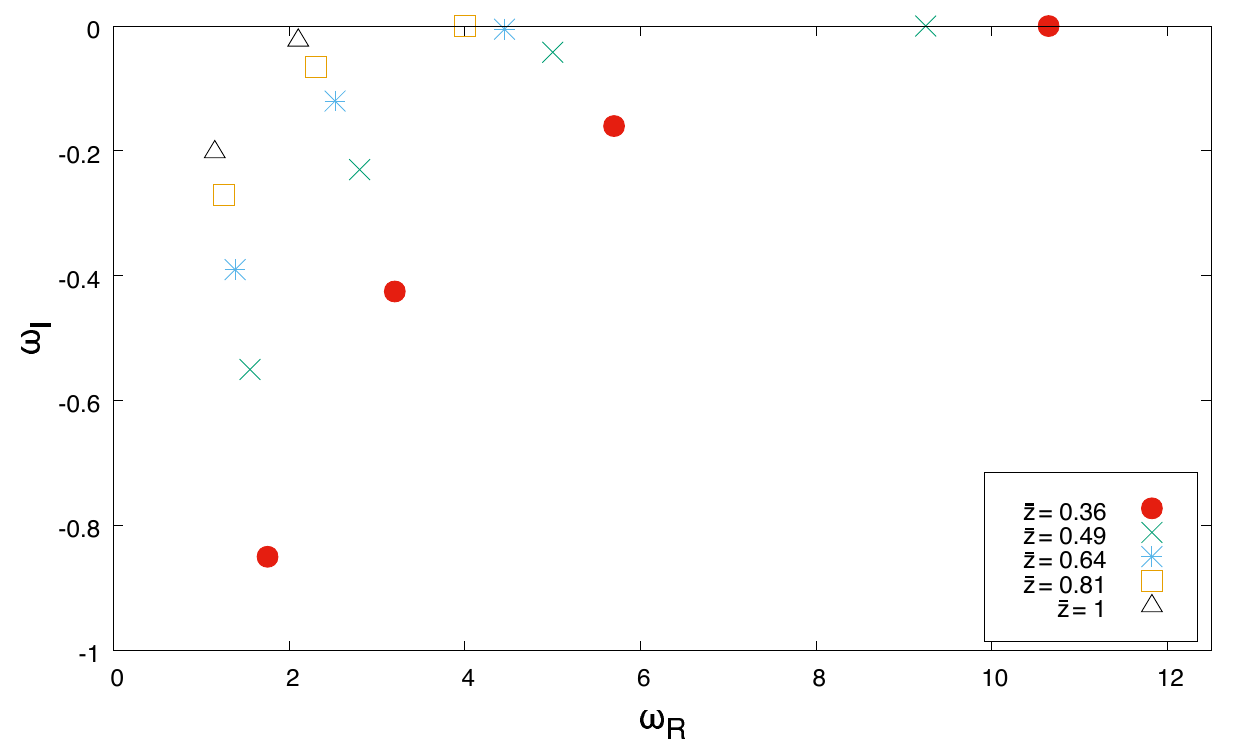}}\nonumber
\end{eqnarray}
\caption{Quasinormal frequencies for $V_+$ (left) and $V_-$ (right)
  for different values of $\bar z$ as indicated in the legend. For
  each $\bar z$, there is a sequence of points, beginning from the smallest
  real frequency, that corresponds to a multipole number $K=1,\, 2,\,
  5,\, 10$ for $V_+$ and $K=2,\, 5,\, 10,\,20$ for $V_-$.}
\label{qnfspin}
\end{figure}

Regarding the multipole number, when $K$, is small the perturbations
decay more rapidly.
One interesting feature displayed in Figs. \ref{spgraphP}
and \ref{spgraphM} is the appearance of long-living nondamped
oscillating modes, the so-called QRMs, for
intermediate $\bar z$ as $K$ grows.
We believe that these QRMs are related to the well-known Klein
paradox. Originally, this paradox appears when studying an electron
hitting a potential barrier~\cite{Klein1929,ItzyksonZuber,Katsnelson}. According to nonrelativistic quantum
mechanics the electron can tunnel the barrier with a damped solution
until a certain penetration distance. However, in relativistic quantum
mechanics, the behavior is different and certainly odd. In fact, when
the barrier's height reaches the mass of the electron $V\sim m_e c^2$,
it becomes almost transparent to it. And even if the barrier becomes
infinite the electron will always tunnel. In our case, it is easy to
see from Figs. \ref{potspin+} and \ref{potspin-} that as $K$ grows the
barrier also grows so that at some point the massless Dirac mode
considered here will borrow enough energy to tunnel and enter the
region with constant potential where it behaves as a free particle.
In Fig. \ref{qnfspin}, we show quasinormal frequencies for several
values of $\bar z$ and $K$. From these graphs, it is clear that QRMs
naturally appear when $\bar z$ is of the same order of $r_+$. This
feature, however, could not be detected using WKB method, the
results of which are shown in Table \ref{dirac_qnm_K_WKB} in the Appendix. This is also clear
in view of the semiclassical character of the WKB approximation, since the Klein paradox
is a quantum effect. In our
case small values of $\sqrt{\bar z}$ compared to the event horizon, {\it i.e.},
far from the Schwarzschild solution, produce a poor WKB
convergence. Nevertheless, we notice that
convergence is much better for large multipole numbers where WKB and
numerical methods produce similar real frequencies.

%%%%%%%%%%%%%%%%%%%%%%%%%%%%%%%%%%%%%%%%%%%%%%%%%%%
\section{Conclusions} \label{final}
%%%%%%%%%%%%%%%%%%%%%%%%%%%%%%%%%%%%%%%%%%%%%%%%%%%
We have considered perturbations on Galileon black holes obtained from
Einstein gravity with a scalar field nonminimally coupled to Einstein
tensor. Vector and fermionic perturbations behave according to the
expectations showing the stability of the model when the $\bar
z$ parameter is positive.

In the case of vector perturbations in both cases, Maxwell and Proca
fields, there is a symmetry by rescaling the spacetime coordinates and
the black hole mass such that quasinormal frequencies obey
Eq.(\ref{rescale}). This fact is also evident from the results
produced by the HH method. In all cases we found no instability under
these vector perturbations.

Regarding Dirac perturbation a new phenomenon, which is similar to
Klein paradox arises; {\it i.e.}, a higher barrier in the potential implies a higher
probability of long-living oscillating modes (QRMs) for intermediate values
of the $\bar z$ parameter and large multipole number $K$. This is a pure
quantum phenomenon that we could only detect by numerically solving
the corresponding Dirac equation. As WKB is a semiclassical approach,
the quasinormal frequencies obtained in this way do not show this
phenomenon. Again no instability under spinorial perturbations has been found so far.

The Galileon black hole model thus shows interesting effects that do not
appear in a simple black hole, and new physics arises, providing new
possible applications in the realm of the AdS/CFT framework. In cosmology
a possible use can only be foreseen in very early phases of the
Universe. Galileon scalar fields describing dark Eenergy are probably
doomed by the effect of these fields in the speed of gravitational
wave propagation, unless some new mechanism occurs.

\acknowledgments

This work was supported by CNPq (Conselho Nacional de Desenvolvimento Cient\'{\i}fico e Tecnol\'ogico), FAPESP (Funda\c c\~ao de Amparo \`a Pesquisa do Estado de S\~ao Paulo) and FAPEMIG (Funda\c c\~ao de Amparo \`a Pesquisa do Estado de Minas Gerais), Brazil.

%%%%%%%%%%%%%%%%%%%%%%%%%%%%%%
\appendix*
\label{appendix}
\section{Results of Horowitz-Hubeny and WKB Methods}
%%%%%%%%%%%%%%%%%%%%%%%%%%%%%%%
\begin{table}[htbp!]
\renewcommand{\arraystretch}{1.3}
\tabcolsep 10pt
\caption{Lowest massless vectorial (Maxwell field) quasinormal
  modes for a Rinaldi black hole with $r_+=0.1$ (left) and $r_+=1$
  (right) produced using the HH method. The multipole number is $\ell=1$. The $*$ signals the
  $\bar z_{cI}$ critical value below which the modes are purely imaginary.}
\label{tabv1}
\begin{tabular}{cccc}
\hline
% &  &  \\
%\cline{1-4}
$\bar{z}$ & $\omega_R $ & $\omega_I$ & $N$ \\
\hline
0.00001&  $\sim 0 $   &$ -2526.8773$ & 40\\
0.00005& $\sim$ 0& $ -528.8917$ & 40\\
0.0001 & $\sim$ 0&  $ -282.4976$ & 40\\
0.000178*&  $\sim 0 $ &    $  -194.0150$& 40 \\
0.0005 & $-27.9461$ & $-67.3439$&40\\
0.001  & $-21.7859$ & $-32.5379$ &50\\
0.005  & $-11.0183$ &  $-5.5403$ &70\\
0.01   & $ -8.3854$  &  $-2.2857$ & 120 \\
%\hline
%\hline
\end{tabular}
\hspace{0.5cm}
\begin{tabular}{cccc}
%\hline
\hline
% &  & \\
%\cline{1-4}
$\bar{z}$ & $\omega_R $ & $\omega_I$ &$N$  \\
\hline
0.001& $\sim0$&$ -252.6877$&30\\
0.005& $\sim0$&$  -52.8892 $&30\\
0.01&  $\sim0$& $ -28.2498 $&30\\
0.0178*&$\sim0$& $-19.4015 $&35\\
0.05& $-2.7946 $   &  $-6.7344$&40 \\
0.1 & $-2.1786 $   &  $-3.2538$ & 60 \\
0.5&  $-1.1018 $   &  $-0.5540$ &70 \\
1&    $-0.8384 $   &  $-0.2286$ & 130\\
%\hline
\hline
\end{tabular}
\end{table}
%%%%%%%%%%%%%%%%%%%%%%%%%%%%%%%%%%%%%%%%%%%%%%%%%%%%%%%%%%
\begin{table}[htbp!]
\renewcommand{\arraystretch}{1.3}
\tabcolsep 10pt

\caption{Lowest massless vectorial (Maxwell field) quasinormal
  modes for a Rinaldi black hole with $r_+=10$ (left) and $r_+=100$
  (right) produced using the HH method. The multipole number is $\ell=1$. The $*$ signals the
  $\bar z_{cI}$ critical value below which the modes are purely imaginary.}
\label{tabv2}
\begin{tabular}{cccc}
%\hline
%\hline
% &  &  \\
%\cline{1-4}
\hspace{0.5cm}
$\bar{z}$ & $\omega_R $ & $\omega_I$ &  $N$ \\
\hline
0.1&  $\sim0 $&$-25.2688$ &40\\
0.5 & $\sim0$ & $-5.2889$  &40 \\
1 &  $\sim 0$ & $-2.8250$&40 \\
1.78*&$\sim 0$& $-1.9401$ & 40  \\
2& $-0.1903$  & $ -1.7498$ & 40 \\
5& $-0.2795$ & $ -0.6734$& 40 \\
10& $-0.2179$ & $ -0.3254$&60 \\
50 &$-0.1102$ & $ -0.0554$&70 \\
150&$-0.0719$ &$  -0.0126$&180 \\
%\hline
\hline
\end{tabular}
\hspace{0.5cm}
\begin{tabular}{cccc}
%\hline
%\hline
%\multicolumn{1}{c}{} &  & \multicolumn{2}{c}{} \\
%\cline{1-4}
$\bar{z}$ & $\omega_R $ & $\omega_I$ &  $N$ \\
\hline
1&  $\sim 0$   &  $  -25.0265$ &30\\
10 & $\sim0$ & $-2.5269$  &30\\
50 &  $\sim0 $   &  $-0.5289$ & 40\\
100 & $\sim 0$   &  $-0.2825$ & 40 \\
178*&  $\sim0$   &  $-0.1940$ & 35\\
180& $-0.0055$ &   $-0.1952$ & 35\\
200&  $-0.0190$  & $-0.1750$ &50\\
500 & $-0.0279$ & $-0.0673$ & 50\\
1000& $-0.0218$ & $-0.0325$ & 50\\
%\hline
\hline
\end{tabular}
\end{table}
%%%%%%%%%%%%%%%%%%%%%%%%%%%%%%%%%%%%%%%%%%%%%%%%%%%%%%%%%%%%%%%%%%%%%%%%

\begin{table}[ht]
\renewcommand{\arraystretch}{1.3}
\tabcolsep 10pt
\caption{Lowest axial massive vectorial (Proca field) quasinormal modes
  for a Rinaldi black hole with $r_+=10$ (left) and $r_+=100$
  (right) produced using the HH method. The multipole number is $\ell=1$. The $*$ signals the
  $\bar z_{cI}$ critical value below which the modes are purely imaginary.}
\label{tabv3}
%%%%%%%%%%%%%%%%%%%%%%%%%%%%%%%%%%%%%%%%%%%%%%%%%
\begin{tabular}{ccccc}
%\hline
%\hline
% &  & & & \\
%\cline{1-5}
$\bar{z}$ & $m$& $\omega_R $ & $\omega_I$ &  $N$ \\
\hline
0.05  &1&  $\sim $   &$ -60.7829 $&80\\
0.122*&1 & $\sim$ &   $ -34.1544 $&90\\
1     &1&  $-2.8317$ &  $-6.0252$ &80\\
0.005 &2 &  $\sim $ &   $-542.4367$ & 50 \\
0.032*&2 &  $\sim$  &   $-128.7811$ &90\\
0.05  &2 &  $-16.8555$ &  $-89.6301$ & 50\\
0.005   & 3 &  $\sim$ &  $-593.9313$&90\\
0.0146* & 3 & $\sim$ &  $-286.0248$ & 90\\
0.05    & 3 & $-34.0195$ & $-101.6521$ & 50\\
%\hline
\hline
\end{tabular}
\hspace{0.25cm}
\begin{tabular}{ccccc}
%\hline
%\hline
% &  & & & \\
%\cline{1-5}
$\bar{z}$ & $m$& $\omega_R $ & $\omega_I$ &  $N$ \\
\hline
0.05  &1&  $\sim $   &$ -603.9213 $&90\\
0.133*&1 & $\sim$ &   $ -315.9499 $&90\\
1     &1&  $-26.2228$ &  $-60.4218$ &50\\
0.005 &2 &  $\sim $ &   $-5420.8095$ & 90 \\
0.033*&2 &  $\sim$  &   $-1256.5409$ &90\\
0.05  &2 &  $-162.93768$ &  $-897.4394$ & 50\\
0.005   & 3 &  $\sim$ &  $-5935.6459$&90\\
0.0148* & 3 & $\sim$ &  $-2843.6852$ & 90\\
0.05    & 3 & $-337.4584$ & $-1017.1671$ & 40\\
%\hline
\hline
\end{tabular}
\end{table}

%\newpage

\begin{table}[ht]
\renewcommand{\arraystretch}{1.3}
\tabcolsep 10pt

\caption{Lowest polar massive vectorial (Proca field) quasinormal
  modes for a Rinaldi black hole with $r_+=10$ (left) and $r_+=100$
  (right) produced using the HH method. The $*$ signals the $\bar z_{cI}$ critical value below which the modes are purely imaginary.}
\label{tabv4}
\begin{tabular}{ccccc}
%\hline
%\hline
% &  & & & \\
%\cline{1-5}
$\bar{z}$ & $m$& $\omega_R $ & $\omega_I$ &  $N$\\
\hline
0.05    &1&  $\sim $ &  $ -4.4840 $&90\\
 1      &1 & $\sim$ &   $ -2.5320 $&50\\
9.6*    &1&  $\sim$ &  $-1.5014$ &50\\
100     &1&  $-0.8276 $ &   $-0.4266$ & 90 \\
150     &1&  $-0.8325$  &   $-0.3495$ &90\\
0.05    & 2 &  $\sim$ &  $-14.8668$ & 120\\
1       & 2 &  $\sim$ &  $-6.5931$  & 50\\
3.31*   & 2 &  $\sim$ &  $-4.7621$  & 50\\
100     & 2 & $-1.6167$ & $-0.8162$ & 90\\
150     & 2 & $-1.6352$ & $-0.6725$ & 90\\
%\hline
\hline
\end{tabular}
\hspace{0.25cm}
\begin{tabular}{ccccc}
%\hline
%\hline
% &  & & & \\
%\cline{1-5}
$\bar{z}$ & $m$& $\omega_R $ & $\omega_I$ &  $N$ \\
\hline
0.05 &1&  $\sim$ &  $-44.8298$ &90\\
1    &1&  $\sim $   &$-25.0032  $&50\\
16.8*&1 & $\sim$ &   $ -10.0507 $&50\\
100  &1 &  $-2.2198$ &   $-3.6808$ & 50 \\
150  &1 &  $-1.8326$  &   $-2.9748$ &50\\
0.05 &2 &  $\sim$ &  $-148.60705$ & 90\\
1    & 2 &  $\sim$ &  $-64.6226$&50\\
4.25*& 2 & $\sim$ &  $-39.8864$ & 50\\
100  & 2 & $-4.2949$ & $-7.1127$ & 50\\
150  & 2 & $-3.5728$ & $-5.7916$ & 50\\
%\hline
\hline
\end{tabular}
\end{table}

\begin{table}[ht]
\renewcommand{\arraystretch}{1.3}
\tabcolsep 10pt
\caption{Dirac quasinormal frequencies with fixed horizon radius
  $r_{+}=1$ computed using the WKB technique.}
\label{dirac_qnm_K_WKB}
  \begin{tabular}{lcccccccc}
   % \hline
    %\hline
    \multirow{3}{*}{$K$} &
      \multicolumn{2}{c}{$\sqrt{\bar z}=1$} &
      \multicolumn{2}{c}{$\sqrt{\bar z}=0.8$} &
      \multicolumn{2}{c}{$\sqrt{\bar z}=0.6$}\\
      \midrule
      & {$\omega_{R}$} & {$\omega_{I}$} & {$\omega_{R}$} & {$\omega_{I}$} & {$\omega_{R}$} & {$\omega_{I}$} &\\
      \hline
    1	& 0.420	& -0.357	& 0.394	& -0.495 	& 0.124	& -1.250	\\
    2	& 0.832	& -0.340	& 0.858	& -0.410		& 0.461	& -0.776 \\
    3	& 1.273	& -0.326	& 1.361	& -0.379		& 1.290	& -0.475 \\
    4	& 1.712	& -0.321	& 1.845	& -0.371		& 1.988	& -0.440 \\
    5   & 2.148 & -0.319	& 2.321	& -0.369		& 2.593	& -0.450 \\
    10  & 4.315 & -0.317    & 4.679 & -0.365        & 5.458 & -0.454 \\
    20  & 8.640 & -0.316    & 9.378 & -0.364        & 10.962 & -0.450 \\
    50  & 21.607& -0.3159   & 23.458 &-0.3631       & 27.439 & -0.449\\
    %\hline
    \hline
  \end{tabular}
\end{table}

%\newpage

\end{document}